\documentclass[a4paper,12pt]{report}
\usepackage[utf8]{inputenc}
\usepackage[top=3cm, left=3cm, right=2cm, bottom=2cm ]{geometry}
\usepackage{graphicx}
\usepackage{times}
\usepackage{pdfpages}
\usepackage{amsmath}
\usepackage{hyperref}
\hypersetup{
    colorlinks=true,
    linkcolor=blue,
    citecolor=blue,
    }
\usepackage[brazil,english]{babel}
\usepackage{comment}
\usepackage{setspace}
\title{}
\author{Ríchard Terra }
\date{}

\begin{document}

\setcounter{page}{0}
\includepdf[pages=1]{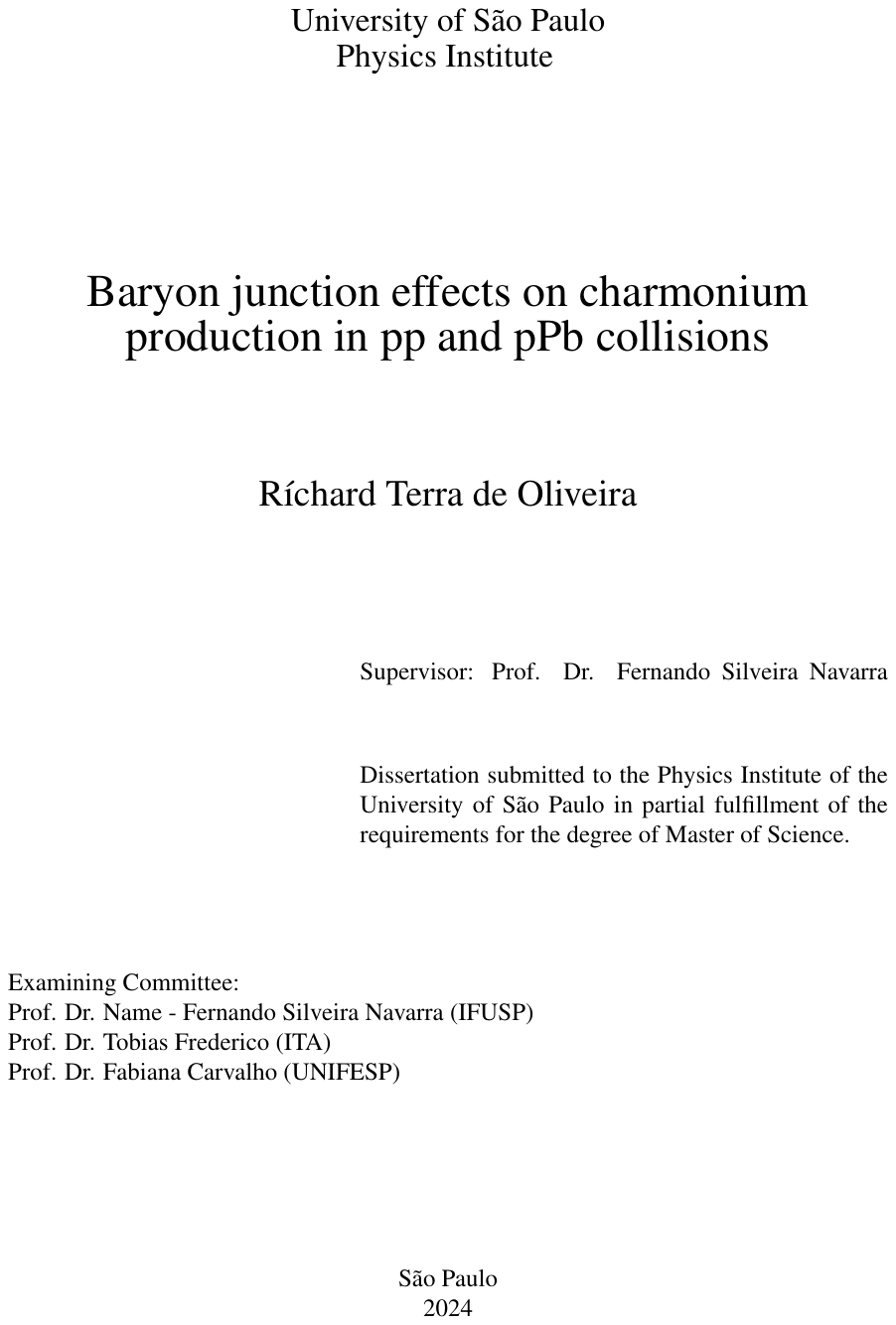}

\thispagestyle{empty}

\singlespacing
\includepdf[pages=1]{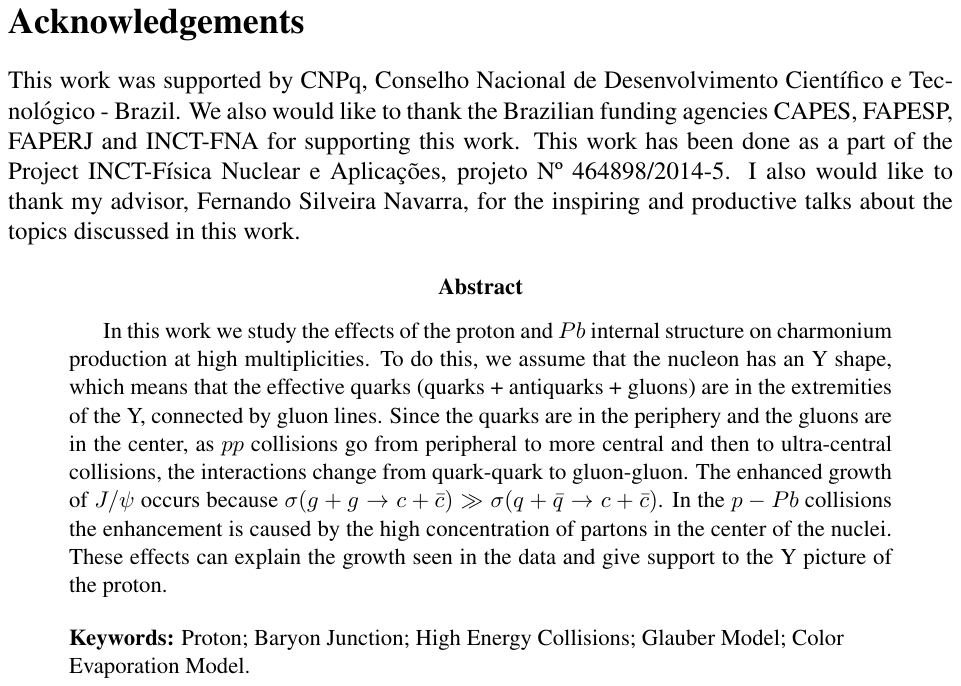}

\selectlanguage{english}

\pagebreak
\thispagestyle{empty}
\onehalfspacing

\tableofcontents
\thispagestyle{empty}

\pagebreak

\chapter{Introduction}

The Large Hadron Collider (LHC) has been providing many interesting ways to study high multiplicity (HM) events and shed light into particle production features in this regime. Conventional meson detection (like charmonium and $D$-mesons) in proton-proton ($pp$) collisions at $\sqrt{s}=7\text{ TeV}$ \cite{alice-psi7,alice-cc} and $\sqrt{s}=13\text{ TeV}$ \cite{alice-psi13}, and proton-lead ($p-Pb$) at $\sqrt{s}=5.02 \text{ TeV}$ \cite{pPjpsi}, have one thing in common:  the meson relative yields grow faster than charged multiplicity density.

The HM events are characterized by small impact parameter configurations, that is, they occur in ultracentral (UC) collisions, and present collective behaviour \cite{alice19}, such as anisotropic flow. Also, in this regime, new features of particle production, such as double-parton scattering and parton saturation, can be responsible for the enhanced growth, and maybe the geometric configuration of the nucleon plays a significant role. 

One interesting static model of the proton, proposed in \cite{kharzeev-bj} and tested in lattice QCD simulations \cite{lattice1,lattice2,lattice3}, leads to the ``Y picture of the proton", where the quarks are in the extremities of an ``Y" like gluon string: the baryon junction. This junction is composed by a large number of gluons, which individually carry small fractions of the proton momentum if compared to the valence quarks (they are in the non-perturbative regime). Because of this difference, gluon flux-tubes of colliding nucleons interact more substantively than the valence quarks, and can also be stopped by soft partons in the midrapidity region. On the other hand, valence quarks will be pulled away and populate the fragmentation region.

Another reason for the existence of the string junction is related to the baryon number conservation. Conventionally, this quantum number should be carried by valence quarks, however experimental data indicate that there is a baryon asymmetry in the midrapidity region \cite{stopping1,stopping2,stopping3,stopping4,stopping5}. Baryon number may be carried not only by valence quarks, but also by sea quarks (quark-antiquark pairs generated by the junction when it is tensioned) \cite{bran22}. This idea  was successfully implemented in analytical models \cite{bopp06}, in the Monte-Carlo event generator HIJING/B \cite{top04}, and explained forward baryon production data. In spite of all the progress, the existence of the baryon junction  still need confirmation \cite{bran22}.

In \cite{deb20} the string junction was used together with the Glauber Model machinery \cite{Glauber-principal}. There the authors used a gaussian ansatz for the gluon and quark anisotropic and inhomogeneous partonic  density  distribution in order to represent the ``Y" shape of the proton. The authors could explain  elliptical flux ($v_2$), measured by the ALICE collaboration in $pp$ collisions at $\sqrt{s}=13 \text{ TeV}$\cite{alice19}. Along this line, we recently adapted the same model to explain successfully the enhanced $J/\psi$ relative yields in UC $pp$ collisions \cite{nosso}.

In this work we are interested in expanding the development we made in \cite{nosso} to explain the enhanced meson production in HM events \cite{alice-psi7,alice-cc,alice-psi13,pPjpsi} by testing the sensibility of the model to different parameter sets and including $pPb$ collisions. Since we will take the average of many collisions and we are not interested in the anisotropic aspects of the collisions, the isotropic gluon distribution proposed in \cite{kubi14} will be used to represent the average configuration of the ``Y" shape over different orientations, i.e., we ``rotate" the ``Y", obtaining one gluon core and a quark corona (``core-corona" model). For charm production, we will use the well known Color Evaporation Model (CEM) \cite{vogt,vogt2}. So, when we go from more peripheral to more central $pp$ collisions, then to UC, we go from quark-quark to gluon-gluon collisions. Since the gluons are much more abundant and $\sigma(gg\rightarrow Q\bar{Q})\gg \sigma(q\bar{q}\rightarrow Q\bar{Q})$, the ``core-corona" structure should explain the enhanced growth in the HM $pp$ collisions, and the higher number of binary collisions should explain the growth seen in $pPb$ collisions.

We start chapter II by presenting some general remarks on the proton structure, and the next by presenting some aspects of HM collisions. In chapter IV we adapt the Glauber model, usually applied to nucleus-nucleus collisions, to $pp$ and $p-Pb$ collisions. In chapter V we combine Glauber outputs (number of participants in a collision and number of binary collisions for a given impact parameter) with CEM, and compare values with experimental data. In the end we make some concluding remarks about the topics discussed in the text.

\chapter{The Proton Structure}

The electron-proton scattering can shed some light into the internal structure of the proton. In this process, the electron exchanges momentum with  the proton by one photon exchange. Depending on the momentum, the aspects of the proton that the process reveals are quite different.

At very low energies, that is, when the wavelength of the virtual photon is much larger than the proton radius, the latter is seen as a point-like particle by the electron and the electrostatic potential is sufficient to understand the process. As the energy increases and the photon wavelength gets closer to the proton radius, the charge and magnetic moment distributions become more important and accessible, and the electrostatic approach is not sufficient. In both cases the struck proton is not broken into hadrons, and the process is said to be elastic.

As the exchanged momentum increases, the wavelength decreases. When the photon wavelength becomes smaller than the proton radius, the process goes from elastic to inelastic scattering. In this case, the virtual photon interacts with the valence quarks and the proton breaks up.  In the very high energy limit, the resolution is such that the proton internal dynamics is completely visible and the sea of quarks and gluons is finally accessed.

In the next sections, the aspects of these different energy regimes will be explored.

\section{Elastic Scattering}

\begin{figure}[h]
    \centering
    \includegraphics[scale=0.6]{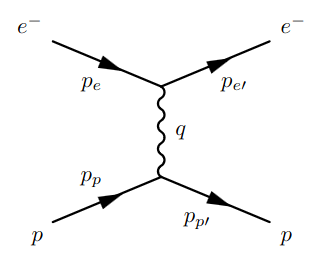}
    \caption{Feynman diagram of $e^-p$ elastic scattering.}
    \label{feynman elastico}
\end{figure}

In very low energy regimes of $e^- p$ scattering, shown if Fig. \ref{feynman elastico}, the energy is such that the proton recoil is negligible compared to its rest mass, and the interaction can be described  by the Coulomb potential. In the limit where the electron is non-relativistic, we have Rutherford scattering, and the differential cross section is given by \cite{thomson}:

\begin{equation}
	\bigg(\frac{d \sigma}{d \Omega}\bigg)_{Ruth} =\frac{\alpha ^2}{16 E^2 \sin^4(\theta/2)}
	\label{ruth}
\end{equation}
where $\alpha=e^2/4\pi$ and $E=p^2/2m_e$ ($m_e$ is the electron mass).

If the proton recoil can still be neglected, but the electron is relativistic, that is, $m_e \ll E \ll m_p$, we have Mott scattering. The respective cross section is given by \cite{thomson}:

\begin{equation}
	\bigg(\frac{d \sigma}{d \Omega}\bigg)_{Mott} =\frac{\alpha ^2}{4 E^2 \sin^4(\theta/2)} \cos^2\bigg(\frac{\theta}{2}\bigg)
	\label{mott}
\end{equation}
where $E=\gamma_e m_e$. 

In both cases, spin and magnetic influences can be neglected, that is, the proton is treated as a spinless nucleus, and as a point-like particle. It is useful to have a way to account for the proton extension and its charge distribution. This is done by introducing form factors. 

Defining the charge density $Q\rho(\mathbf{r}\prime)$, where $Q$ is the total charge and $\rho(\mathbf{r}\prime)$ is the normalized charge distribution, the interaction can be described by means of the potential:

\begin{equation}
	V(\mathbf{r})=\int \frac{Q\rho(\mathbf{r}\prime) }{4\pi \|\mathbf{r}-\mathbf{r\prime}\|}d^3 \mathbf{r\prime}
\end{equation}

Considering $e$ as the incoming electron, $e\prime$ as the outgoing electron and that their wave functions are  plane waves (Born approximation), the matrix element can be written as:

\begin{eqnarray}
\mathcal{M} &=& \left \langle \psi_{e\prime} | V(\mathbf{r}) | \psi_e \right \rangle = \int e^{-i\mathbf{p_{e\prime}}\cdot\mathbf{r}} V(\mathbf{r)}e^{i\mathbf{p_e}\cdot\mathbf{r}} d^3 \mathbf{r} \\
           &=& \int d^3 \mathbf{R} e^{i\mathbf{q}\cdot \mathbf{R}} \frac{Q}{4\pi |\mathbf{R}|} \int d^3 \mathbf{r\prime} \rho(\mathbf{r\prime}) e^{i\mathbf{q}\cdot \mathbf{r\prime}}\\
           &=& \mathcal{M}^{pl}F(\mathbf{q}^2)
\end{eqnarray}
with $\mathbf{q}=\mathbf{p_e}-\mathbf{p_{e\prime}}$, $\mathbf{R}=\mathbf{r}-\mathbf{r\prime}$, $\mathbf{M}^{pl}$ is the point-like proton matrix element and $F(\mathbf{q}^2)$ is the electric form factor. The last one can be interpreted as the Fourier transform of the charge distribution. Since the matrix element still depends on point-like approach, the differential cross section for Mott scattering is changed to:

\begin{equation}
	\bigg(\frac{d \sigma}{d \Omega}\bigg)_{Mott} =\frac{\alpha ^2}{4 E^2 \sin^4(\theta/2)} \cos^2\bigg(\frac{\theta}{2}\bigg) |F(\mathbf{q}^2)|^2
	\label{mottmod}
\end{equation} In other words, the charge distribution inside the proton can be indirectly measured through the differential cross section measurement in the low energy limit. 

Coming from the very low energy limit to the low energy, that is, when the exchange momenta $|\mathbf{q}|$ gets closer to proton mass, the recoil and spin interactions become more important. If the proton is still treated as a point-like particle, but as a spin $1/2$ particle and with recoil, the differential cross section is given by \cite{thomson}:

\begin{equation}
	\bigg(\frac{d \sigma}{d \Omega}\bigg) =\frac{\alpha ^2}{4 E^2_e \sin^4(\theta/2)}\frac{E_{e\prime}}{E_e} \bigg( \cos^2 \frac{\theta}{2} +\frac{Q^2}{2m_p^2} \sin^2\frac{\theta}{2} \bigg)
\end{equation}
with $Q^2=-q^2=2m_p(E_e-E_{e\prime})$ as the virtuality.

Like before, in the most general case (proton not treated as a point-like particle) the internal structure can be accounted for by introducing the electric form factor $G_E(Q^2)$ and magnetic form factor $G_M(Q^2)$. The differential cross section can be written as the Rosenbluth formula:

\begin{equation}
	\bigg(\frac{d \sigma}{d \Omega}\bigg) =\frac{\alpha ^2}{4 E^2_e \sin^4(\theta/2)}\frac{E_{e\prime}}{E_e} \bigg( \frac{G_E^2+\tau G_M^2}{1+\tau}\cos^2 \frac{\theta}{2} + 2\tau G_M^2 \sin^2\frac{\theta}{2} \bigg)
	\label{rosenbluth}
\end{equation}
with $\tau=Q^2/4m_p^2$.

Differently from the $F(\mathbf{q}^2)$, form factors can not be directly understood as Fourier transforms of charge and magnetic momenta distribution. They are functions of the four-momentum $Q^2$, not the three-momentum $q^2$. Explicitly:

\begin{eqnarray}
Q^2 &=& -q^2=\mathbf{q}^2-(E_e-E_{e\prime}) \\
           &\Rightarrow &  Q^2 \bigg(1+ \frac{Q^2}{4m_p}\bigg)= \mathbf{q}^2  
\end{eqnarray}
So $Q^2\approx \mathbf{q}^2$ if $Q^2\ll 4 m_p^2$, and the form factors can be interpreted as:

\begin{equation}
	G_E(Q^2)\approx G_E(\mathbf{q}^2) = \int e^{i\mathbf{q}\cdot \mathbf{r}} \rho(\mathbf{r}) d^3\mathbf{r}
\end{equation}

\begin{equation}
	G_M(Q^2)\approx G_M(\mathbf{q}^2) = \int e^{i\mathbf{q}\cdot \mathbf{r}} \mu(\mathbf{r}) d^3\mathbf{r}
\end{equation}

The measurement of the electric form factor, the one which we are mainly interested in, leads to the exponential charge distribution of \cite{thomson}

\begin{equation}
	\rho(\mathbf{r})\approx \rho_0 e^{-r/a}
\end{equation}
with $a\approx0.24 \text{ fm}$, what corresponds to a proton root-mean square charge radius around $0.8\text{ fm}$.

Once the electric form factor is measured through the differential cross section measurement of Eq.(\ref{rosenbluth}), the proton charge radius can be calculated in electron-proton collisions as \cite{puzzle1}:

\begin{equation}
	\left \langle r_p^2 \right \rangle = -\frac{6}{G_E(0)} \frac{dG_E(Q^2)}{dQ^2} \bigg|_{Q^2=0}
\end{equation}
In 2010, the CODATA-2010 \cite{CODATA10} value for the charge radius of the proton, obtained from $ep$ scattering, was shown to be $r_p= 0.895 (18) \text{ fm}$ . 

Another way to obtain the proton charge radius is through hydrogen spectroscopy. In this approach the electron energy is shifted when it moves inside the proton, experiencing an screening effect. The CODATA-2010 value for the radius in this method was shown to be $r_p=0.8758 (77) \text{ fm}$ and the final proton charge radius was $r_p=0.8775(51) \text{ fm}$.

In 2010 and 2013, the CREMA collaboration \cite{CREMA2010,CREMA2013} found the charge radius through the muonic hydrogen ($\mu  H$) spectroscopy method. In this experiment, the electron from the hydrogen atom is replaced by one muon. Since the muon is much heavier than the electron, the Bohr radius of hydrogen decreases and the sensitivity to the proton size increases. The extraordinarily precise results found were $r_p= 0.84184 (67) \text{ fm}$ and $r_p= 0.84087 (39) \text{ fm}$. 

Since values from CREMA and CODATA were around $7\sigma$ of difference, and around $4\%$, this discrepancy became  what is known as ``the charge radius puzzle". This puzzle is not completely solved until now. Some results from lattice and lepton-muon scattering are discussed in \cite{puzzle1,puzzle2}.

\section{Deep Inelastic Scattering and the Quark-Parton Model}

\begin{figure}[h]
    \centering
    \includegraphics[scale=0.6]{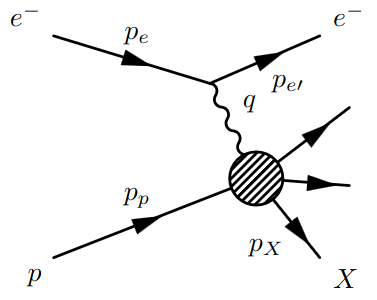}
    \caption{Feynman diagram of $e^-p$ deep inelastic scattering.}
    \label{feynman inelastico}
\end{figure}

Moving away from the elastic cross section by increasing the energy, the study of the proton structure must change. While the virtuality ($Q^2$) increases, the exchanged photon starts to interact with the quarks that compose the proton, which is broken. The process is now represented by $e^- p \rightarrow e^- X$ in Fig. \ref{feynman inelastico}, where $X$ is the hadronic final state.

For the deep inelastic scattering, the double differential cross section can be written as \cite{thomson}:

\begin{equation}
	\frac{d^2 \sigma}{dx dQ^2}= \frac{4\pi \alpha^2}{Q^2} \bigg[(1-y) \frac{F_2(x,Q^2)}{x} + y^2 F_1(x,Q^2) \bigg]
	\label{doubdifincs}
\end{equation}

In this approach, the form factors of Eq.(\ref{rosenbluth}) are absorbed into the structure functions $F_1(x,Q^2)$ and $F_2(x,Q^2)$, which can not be interpreted as the Fourier transforms of charge and magnetic momentum distribution since there is an $x$ and $Q^2$  dependence. The values $x$ and $y$ are Bjorken-$x$ and inelasticity, which take values in the range $[0, 1]$, and are defined as:

\begin{equation}
	x=\frac{Q^2}{2 p_p \cdot q}=\frac{Q^2}{Q^2+p_X ^2 - m_p^2}
	\label{bjorkenx}
\end{equation}

\begin{equation}
	y=\frac{p_p \cdot q}{p_p \cdot p_e}= 1- \frac{E_{e\prime}}{E_e}
	\label{inelasticity}
\end{equation}

The structure functions are not independent from each other in the regime with $Q^2$ greater than a few GeV, but they are almost independent of $Q^2$ (Bjorken-scaling), and are related by the Callan-Gross relation:

\begin{equation}
	F_2(x)=2x F_1(x)
	\label{callan-gross}
\end{equation}

The quark-parton model was proposed by Feynman. In this model, the deep inelastic $e^-p$ interaction is interpreted  as electron-quark elastic interactions, with quarks as free particles. 

This model is applied in the infinite-momentum frame, the one in which the proton energy is much bigger than the respective rest mass. The advantage of this approach is that the valence quarks can be considered to be moving in the proton beam line, and each quark carries a fraction $x$ of proton momentum. This $x$ value is the same Bjorken-x, which now is interpreted as the fraction of proton momentum carried by the parton.

Since the valence quarks will interact with each other by gluon exchanges, the momentum distribution inside the proton will be given by the parton distribution functions (PDFs). These functions represent the probabilities of a given parton of flavor $i$ to have a momentum fraction in the range $x$ and $x+\delta x$. These distributions are not known directly from theory because there are non-perturbative aspects involved, but they are experimentally obtained.

In the end, the double differential inelastic cross section of Eq.(\ref{doubdifincs}) can be rewritten as \cite{thomson}:

\begin{equation}
	\frac{d^2 \sigma}{dx dQ^2}= \frac{4\pi \alpha^2}{Q^4} \bigg[(1-y) + \frac{y^2}{2}\bigg] \sum_i Q_i^2 f_i^2(x)
	\label{doubdifincsfinal}
\end{equation}
where $Q_i$ is the flavour charge and $f_i$ is the respective PDF.

The PDFs must reflect some internal structure features of the proton. The proton can be considered as being composed by $uud$ quarks. These valence quarks interact by exchanging gluons, which can fluctuate into virtual $q\bar{q}$ pairs. These produced quarks are said to be ``sea quarks" and are mainly produced in the low $x$ regime. In the end, the $e^- p$ scattering must include the interactions between electron and sea quarks. 

Once there are $2$ valence quarks up and $1$  valence quark down, the PDFs must be such that \cite{peskin}:

\begin{equation}
	\int_0^1 dx [f_u(x)-f_{\bar{u}}(x)]=2
\end{equation}
and

\begin{equation}
	\int_0^1 dx [f_d(x)-f_{\bar{d}}(x)]=1
\end{equation}
Also, because the total momentum carried by the partons must be the total momentum of the proton, we must have \cite{peskin}

\begin{equation}
	\int_0^1 dx \text{ }x[f_u(x)+f_{\bar{u}}(x)+f_d(x)+f_{\bar{d}}(x)+g_g(x)]=1
\end{equation}
and the gluons contribute with half of the total value, that is, they carry half of the proton momentum!

Although the PDFs were assumed to be only dependent on $x$, at very low $x$ and at $x\approx1$ the structure functions vary strongly with $Q^2$. In the first case the functions get higher, and in the second, get lower, while $Q^2$ increases.  The evolution of PDFs in $Q^2$ are made through the DGLAP equations developed in \cite{peskin,dglap}.

In summary, coming from elastic scattering to deep inelastic scattering, the proton goes from one symmetric charge distribution configuration to a more complex distribution, which implies an even more complex charge distribution inside it. Also, coming from $x\approx1$ regime to the low $x$ regime, the proton goes from a three valence quark composition to a complex internal matter distribution generated by gluon fluctuations. In the end, the gluon body starts to play an important role in the internal proton dynamics.

\section{The Gravitational Form Factors (GFFs)}

The electromagnetic properties of the proton shed light into many interesting aspects of the internal structure of the proton. However, there are particles inside it, like gluons, that do not interact electromagnetically, but are important pieces of the nucleon. For example, the proton mass, which is around $1\text{ GeV}$, can not be only composed by valence quarks, which present masses around few MeV, generated by the Higgs boson. So, although gluons do not have rest mass, part of the proton mass can be generated by their influence in the energy momentum tensor. Also, part of the mass can be generated by the trace anomaly \cite{TA1,TA2,TA3}.

\begin{figure}[h]
    \centering
    \includegraphics[scale=0.4]{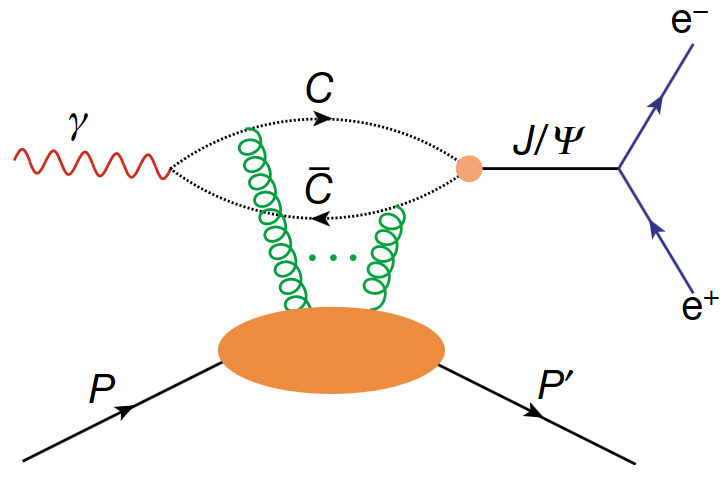}
    \caption{Feynman diagram of the $J/\psi$ photopropduction and its interaction with the gluon composition of the proton \cite{nature}.}
    \label{feynman gluon}
\end{figure}

The gluon distribution is not easily accessed because gluons do not have charge. The gluon content can be seen through the $J/\psi$ photoproduction process near threshold, at the Jefferson Lab. In this process the photon  fluctuates into $c\bar{c}$ pair and exchanges gluons with the struck proton before decaying into $J/\psi$, and then to $e^-e^+$. The Feynman diagram of the process is shown in Fig. \ref{feynman gluon}.

The gravitational form factors for a spin $1/2$ hadron can be obtained from the energy-momentum tensor matrix elements as \cite{kharzeev GFFs,Yoshi GFFs}:

\begin{eqnarray}
\left \langle P\prime | T_{q,g}^{\mu \nu} | P\right \rangle &=& \bar{u}(P\prime)  \bigg[\big(A_{q,g}(t)+B_{q,g}(t) \big)\gamma^{(^\mu \bar{P} ^\nu)} - \frac{\bar{P}^\mu \bar{P}^\nu}{M} B_{q,g}(t)
 \nonumber \\
                     &+&  C_{q,g}(t) \frac{\Delta^\mu \Delta^\nu - g^{\mu\nu}\Delta^2}{M} + \bar{C}_{q,g}(t) M \eta^{\mu \nu} \bigg] u(P)
\end{eqnarray}
where $A_{q,g}(t)$ is related to the quark and gluon momenta, $J_{q,g}(t)=1/2 \big(A_{q,g}(t)+B_{q,g}(t) \big)$ is related to the angular momentum and $D_{q,g}=4 C_{q,g}(t)$ is related to the pressure and shear forces.

The photoproduction differential cross section as a function of $t$ Mandelstam variable, for different photon energies ($9.1 \text{ GeV} \le E_{\gamma} \le10.6\text{ GeV} $), was measured in \cite{nature}. The gravitational form factors were extracted by fitting the data through the holographic QCD approach and the generalized parton distribution (GPD) approach. Both of them depend explicitly on form factors. After finding them, the mass radius ($r_m$) and scalar radius ($r_s$) can be calculated as:

\begin{equation}
	\left \langle r_m ^2 \right \rangle =  \frac{6}{A_{g}(0)} \frac{d A_{g}(t)}{dt} \bigg |_{t=0} - \frac{6}{A_g(0)} \frac{C_g(0)}{M_p^2}
\end{equation}
and
\begin{equation}
	\left \langle r_s ^2 \right \rangle =  \frac{6}{A_{g}(0)} \frac{d A_{s}(t)}{dt} \bigg |_{t=0} - \frac{18}{A_g(0)} \frac{C_g(0)}{M_p^2}
\end{equation}
The mass radius makes reference to the tensor gluon field structure, which provides most of the proton mass, and the scalar radius makes reference to a confining scalar gluon density. In Table \ref{naturerad} the radius obtained from \cite{nature} through the different fits and the lattice value \cite{latticerm} are shown. We can notice that the mass radius is much smaller than the charge radius, which is around $0.8 \text{ fm}$, and the scalar radius goes far beyond it.

The scalar and mass radius of the proton gives important information about the size of its gluon component. They suggest that it is composed by three layers: one internal gluon distribution, one thin quark layer and one external gluon layer. There still is one interesting question that arises: how are the gluons spatially distributed?

\begin{table}[h!]
\caption{Mass radius of the proton obtained in \cite{nature} through different fits of differential cross section and lattice result\cite{latticerm}.}
\label{naturerad}
\centering
    \begin{tabular}{| c | c | c |}
    \hline
    Models & $\sqrt{\left\langle r_m^2 \right\rangle}\text{ } (\text{fm})$ & $\sqrt{\left\langle r_s^2 \right\rangle}\text{ } (\text{fm})$ \\ \hline
    Holographic QCD & $0.755 \pm 0.035$ & $1.069 \pm 0.056$ \\ \hline
    GPD             & $0.472 \pm 0.042$ & $0.695 \pm 0.071$\\ \hline
    Lattice         & $0.7464 \pm 0.025$ & $1.073 \pm 0.066$\\ \hline
    \end{tabular}
\end{table}

\section{The Baryon Junction}

\begin{figure}[h]
    \centering
    \includegraphics[scale=0.6]{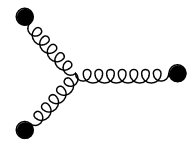}
    \caption{Baryon junction.}
    \label{Baryon junction}
\end{figure}

In \cite{kharzeev-bj} we find a really interesting spatial distribution of gluons inside the proton. According to  this picture, the valence quarks are in the extremities of an ``Y" shaped gluon string, called the ``Baryon-Junction". It is shown in Fig. \ref{Baryon junction}.

In this picture, the string gluons carry, individually, a small fraction of  the proton momentum. So, in high energy collisions, the gluons interact more substantively, stopping in the midrapidity region, while the valence quarks pass through each other and go to the fragmentation region. While the baryon junction stops and the valence quarks are pulled away, the string is tensioned, and $q\bar{q}$ pairs are created.

The produced pairs must present some characteristics \cite{bran22}: they present low transverse momentum because they come from soft partons process; since the junction is flavour blind, the internal content does not have to be the same of the colliding baryons;  they will be accompanied by many pions; the baryon-junction stopping is characterized by the exponential drop of the cross section with the rapidity loss variable. Right after the collision, the stopped junction generates baryons composed by sea quarks, which can differ from projectile or target nucleons, and the number of baryons in central rapidity region exceeds the number of antibaryons \cite{baryon-antibaryon}.

The initial motivation that led to the baryon-junction picture was the guiding question: ``Can Gluons Trace Baryon Number?". This is the title of the paper that first proposed the model \cite{kharzeev-bj}. Since valence quarks carry color, flavour, electric charge and isospin, it is natural to state that they also carry the baryon number, but this assumption is not dictated by QCD. So, part of the number can be carried by the junction.

The baryon-junction manifestations are expected to be seen in deep inelastic $e+p/AU$ scattering at the future electron-ion collider (EIC) \cite{bran22}, in inclusive ultra-peripheral collisions at RHIC \cite{bran22} and in ultra-central $pp$ collisions. In the last one this happens because the gluons are mainly concentrated in the central region of the protons, generating high multiplicity collisions. The aspects of this regime will be better explored in the next chapter.

\chapter{High Multiplicity $pp$ Collisions}

    High multiplicity $pp$ events present some interesting characteristics. Since they occur for small impact parameters (UC collisions), and gluons are localized in the center, the interactions involve mainly gluons and sea quarks. There are also collective behaviour features that suggest formation of quark-gluon plasma.

The collective behaviour can be generically described in terms of correlated particle production. Considering $P(\mathbf{p}_i)$ as the probability of  producing a particle with momentum $\mathbf{p}_i$, there is correlation behaviour if $P(\mathbf{p}_1,\mathbf{p}_2)\ne P(\mathbf{p}_1)P(\mathbf{p}_2)$ \cite{qgp1}. That is, the probabilities can not be separated. The correlation is usually measured and plotted by means of relative azimuthal angle $\Delta \phi$ in the transverse plane, and relative longitudinal pseudorapidity $\Delta \eta$.

\begin{figure}[h!]
\centering
\begin{tabular}{cc}
    \includegraphics[scale=0.6]{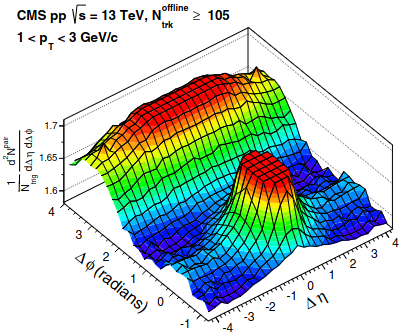}& \includegraphics[scale=0.6]{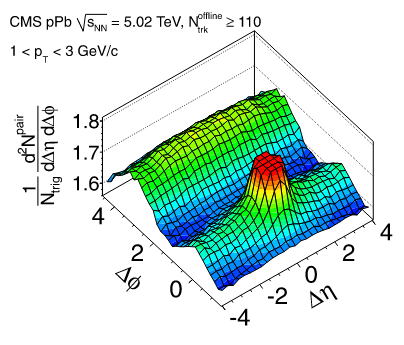} \\
    (a) &  (b)\\
    
     \includegraphics[scale=0.6]{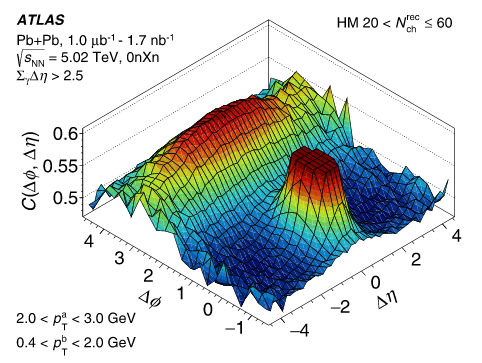}&  \includegraphics[scale=0.45]{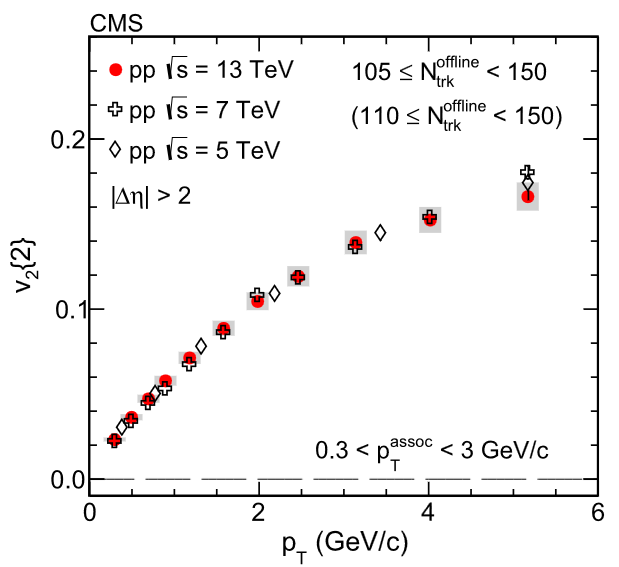} \\
  (c)&    (d)
\end{tabular}
    \caption{a) pp b) pPb and c) PbPb  correlation measurements \cite{correlpp,correlpPb,correlPbPb}, and elliptical flux coefficient $v_2$ \cite{v2pp}.}
\label{correl}
\end{figure}

The angle and pseudorapidity distribution can be written as 

\begin{equation}
	\frac{d^2N}{d  \Delta \phi d \Delta\eta} \propto 1 +2 \sum_n v_n(p_T,\Delta\eta) \cos(n\Delta \phi)
	\label{ptetadistrib}
\end{equation}
The pair correlation measurements for $pp$, $pPb$ and $PbPb$ collisions are shown in Fig. \ref{correl} a), b), c). In all the cases, there are ridges in $\Delta \phi\approx0$ and $\Delta \phi\approx\pi$, with peaks in $\Delta \eta\approx0$. Since there is QGP formation in $PbPb$ collisions at $\sqrt{s}= 5.02 \text{ TeV}$ and $pp$, $pPb$ high multiplicity collisions present similar behaviour in the correlation measurements if compared to the $PbPb$ case, the QGP formation seems to occur also in small systems.

In Eq.(\ref{ptetadistrib}) the $v_n$ coefficients contains important information about the collective behaviour. The first four ($v_1$,$v_2$,$v_3$,$v_4$), also known as directed, elliptic, triangular and quadrangular flow coefficients, reflect pressure gradients along the transverse plane. The $p_T$ dependence of elliptical flux for high multiplicity $pp$ collisions, for pair correlation, is shown in Fig. \ref{correl} d). 
\pagebreak

\begin{figure}
\centering
\begin{tabular}{c}
    \includegraphics[scale=0.65]{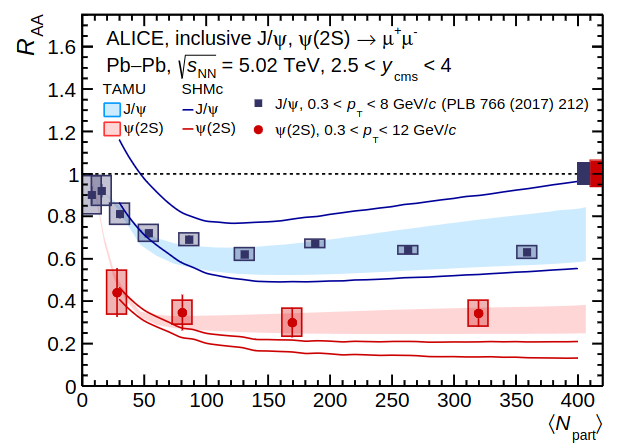}\\ 
    \includegraphics[scale=0.45]{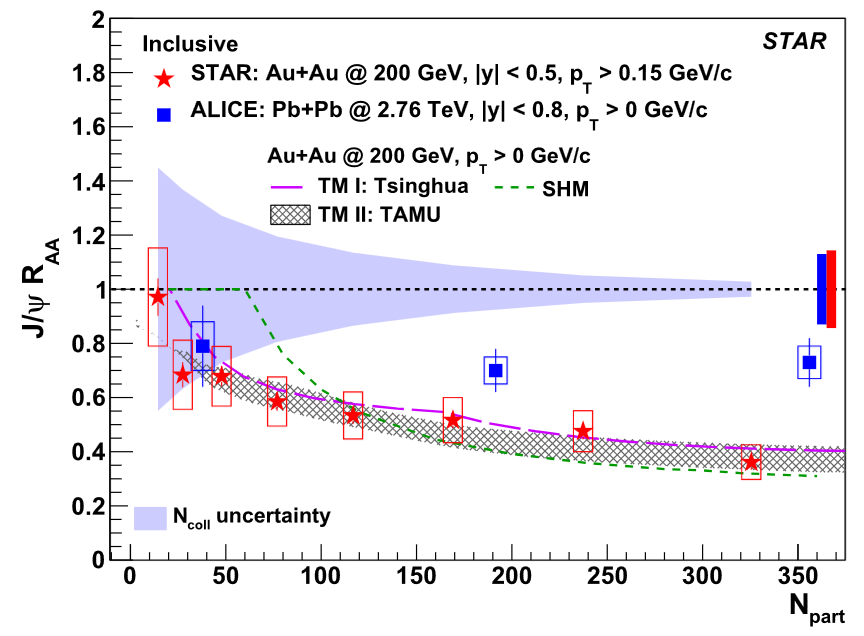} \\

\end{tabular}
    \caption{Charmonium suppression in $PbPb$  \cite{RAA,RAA2,RAA3,RAA4} at LHC and $AuAu$ collisions at RHIC \cite{RAA5}.}
\label{sup reg}
\end{figure}

Another interesting QGP signature is $J/\psi$ suppression. For heavy-ion collisions, when the medium temperature is higher than the dissociation temperature of the meson, this one is suppressed due to screening. The nuclear modification factor as a function of $N_{part}$ (number of nucleons that participate in the collisions) for $PbPb$ collisions at the LHC is show in Fig. \ref{sup reg}. As can be noted, as collisions get more central ($N_{part}$ increases), $J/\psi$ and $\psi(2S)$ mesons are suppressed. In Fig. \ref{sup reg} we can see that for lower energies the suppression is more intense than in higher energies, which means that quarkonia is regenerated. In principle, if there is QGP formation in $pp$ collisions, an analogue pattern of suppression is expected.

\begin{figure}
\centering
\begin{tabular}{c}
    \includegraphics[scale=0.65]{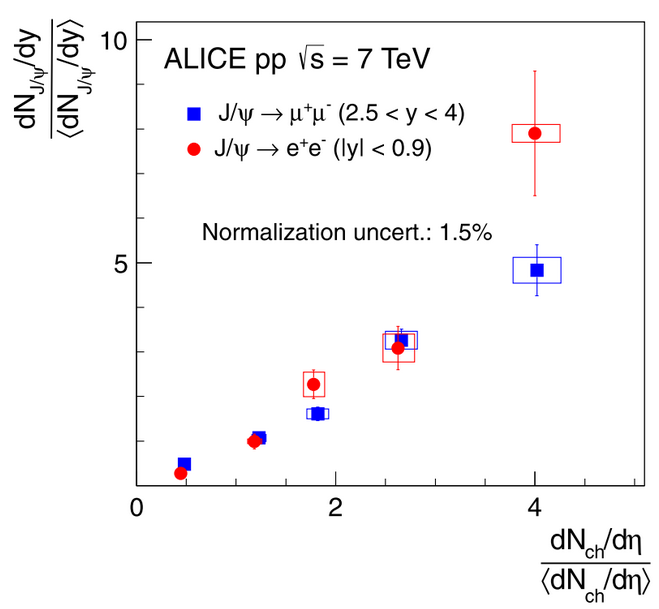}\\ 
    (a)\\
    \includegraphics[scale=0.6]{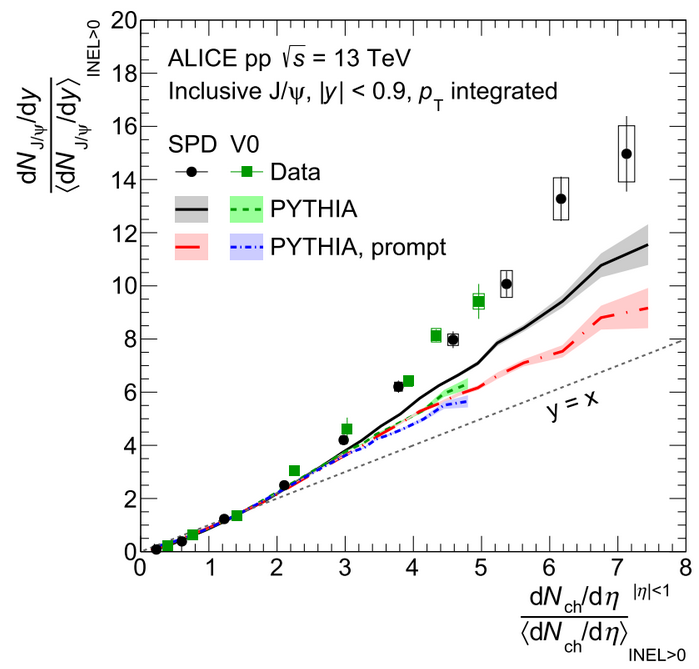} \\
    (b)\\
\end{tabular}
    \caption{$J/\psi$ relative yields per charged multiplicity at a) $\sqrt{s}=7\text{ TeV}$ and  b) $\sqrt{s}=13\text{ TeV}$. The experimental data are from \cite{alice-psi7,alice-psi13}.}
\label{jpsiALICE}
\end{figure}

The $J/\psi$ relative yields per relative charged multiplicity at midrapidity in $pp$ collisions were measured by ALICE collaboration at $\sqrt{s}=7\text{ TeV}$ and $\sqrt{s}=13\text{ TeV}$. The results are shown in Fig. \ref{jpsiALICE}. The data present an anomalous behaviour. In the low multiplicity region (peripheral collisions) they grow linearly as expected, but in the high multiplicity region (UC collisions), production is enhanced. Since gluons are mainly concentrated in central region of the proton, this growth may be explained by their presence. In this work, as will be discussed latter, we will try to interpret the data on $J/\psi$ production in UC region as a manifestation of the baryon junction. 

The presence of QGP in ultra-central $pp$ collisions is still an open question and some general aspects were discussed in \cite{qgp1,qgp2,qgpjpsi}.

\chapter{Glauber Model}

Firstly proposed to deal with nucleus-nucleus collisions, the Glauber model provides an interesting way of introducing geometrical variables in high energy collisions. For a given nucleon number density distribution, it returns the number of binary collisions between nucleons ($N_{coll}$) and the number of nucleons that participate in the collisions ($N_{part}$) at a given impact parameter.  Because of the density input, the model can shed some light in the internal structure of the entities involved in the collisions.

The nucleon motion and interactions during a collision can be really difficult to predict, but some approximations can reduce the treatment to simple integrals involving the density. Because of the high center of mass energy, the nucleons carry sufficient momentum to be mainly undeflected while the nuclei pass through each other. Also, it is assumed that the nucleus sizes are much bigger than the range of the nucleon-nucleon forces \cite{Glauber-principal}. In the next section, the formalism of the model will be developed by considering the approximations above.

\section{Formalism}

\begin{figure}[h]
    \centering
    \includegraphics[scale=0.8]{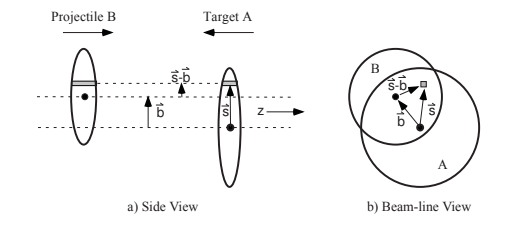}
    \caption{Glauber model geometry \cite{Glauber-principal}.}
    \label{glauber model geometry}
\end{figure}

The geometrical formulation of the collision is shown in Fig. \ref{glauber model geometry}. $\vec{b}$ is the impact parameter, $\vec{s}$ is the position of a given nucleon flux tube, $A$ is the target and $B$ is the projectile. The same two letters are used to indicate the respective number of nucleons.

The probability per unit of inverse area of finding a given nucleon in the target flux tube, or in the projectile flux tube, in a position $\vec{s}$, is, respectively,

\begin{equation}
	T_A (\vec{s})= \int \rho_A (\vec{s},z_A) dz_A
	\label{TA0}
\end{equation} and 

\begin{equation}
	T_B (\vec{s})= \int \rho_B (\vec{s},z_B)dz_B
    \label{TB0}
\end{equation} 
With $\rho_{A,B}$ as the respective densities. The same probabilities per nucleon are

\begin{equation}
	\hat{T}_A (\vec{s})= \int \frac{\rho_A (\vec{s},z_A)}{A} dz_A
	\label{T}
\end{equation} and 

\begin{equation}
	\hat{T}_B (\vec{s})= \int \frac{\rho_B (\vec{s},z_B)}{B}dz_B
    \label{TB}
\end{equation} 

The probability per unit of area for the interaction of a nucleon from $A$ and another from $B$ to occur in a overlap area can be calculated as  

\begin{equation}
    \hat{T}_{AB}(\vec{b})=\int \hat{T}_A (\vec{s}) \hat{T}_B(\vec{s}-\vec{b}) d^2s
    \label{overlap}
\end{equation} 

The product $\hat{T}_{AB}(\vec{b})\sigma_{inel}^{NN}$ (which has no units) represents the probability of an interaction ($\sigma_{inel}^{NN}$ is the nucleon-nucleon inelastic cross section, which depends on the center of mass energy). Along this line, the probability of having $n$ interactions in the collision is

\begin{equation}
    P(n,\vec{b})={AB \choose n} \big( \hat{T}_{AB}(\vec{b}) \sigma_{inel}^{NN}  \big)^n \big(1- \hat{T}_{AB}(\vec{b}) \sigma_{inel}^{NN}  \big)^{AB-n}
    \label{P(n,b)}
\end{equation} 
The first term is the  combination of $n$ collisions with $AB$ possibilities, the second represents the probability of $n$ interactions, and the third represents $AB-n$ missed interactions.

Considering non-polarized nucleons, $\Vec{b}\rightarrow b$, and the probability treatment leads to the calculation of the total number of binary collisions

\begin{equation}
    N_{coll}=\sum_{n=1}^{AB}nP(n,b)=AB\hat{T}_{AB}(b)\sigma_{inel}^{NN}
    \label{Ncoll}
\end{equation} 
which is the probability of one interaction multiplied by the possible number of interactions.

The number of participants (also called wounded nucleons) can be calculated as \cite{Npart1,Npart2}

\begin{eqnarray}
N_{part}(\vec{b}) &=& A\int \hat{T}_A(\vec{s}) \big\{ 1-\big(1- \hat{T}_{B}(\vec{s}-\vec{b}) \sigma_{inel}^{NN}  \big)^{B}\big\}d^2s \\
                     &+& B\int \hat{T}_B(\vec{s}-\vec{b}) \big\{ 1-\big(1- \hat{T}_{A}(\vec{s}) \sigma_{inel}^{NN}  \big)^{A}\big\}d^2s
\label{Npart}
\end{eqnarray}

Sometimes, mainly because the calculation of $N_{part}$ is demanding, it is useful to estimate it according to the scaling behaviour of Fig. 7 of \cite{Glauber-principal} as:

\begin{equation}
    N_{part}(b)\propto N_{coll}^{3/4}(b)
    \label{Npart estimado}
\end{equation}

\section{Pb-Pb Collisions}

The heavy ion nucleon density shape is usually assumed to be the Woods-Saxon 

\begin{equation}
    \rho(r)=\rho_0\frac{1+w(r/R)^2}{1+\exp\big(\frac{r-R}{a}\big)}
    \label{nuclear charge density}
\end{equation}  where $\rho_0$ is the nucleon density in the center of the nucleus, $R$ is the radius of the nucleus, $w$ represents deviations from the sphere format, and $a$ is the skin depth. The normalization constant $\rho_0$ must be such that 

\begin{equation}
    \int d^3 \mathbf{r} \rho_A(\mathbf{r})=A
    \label{A normalization}
\end{equation}

\begin{figure}[h]
    \centering
    \includegraphics[scale=0.5]{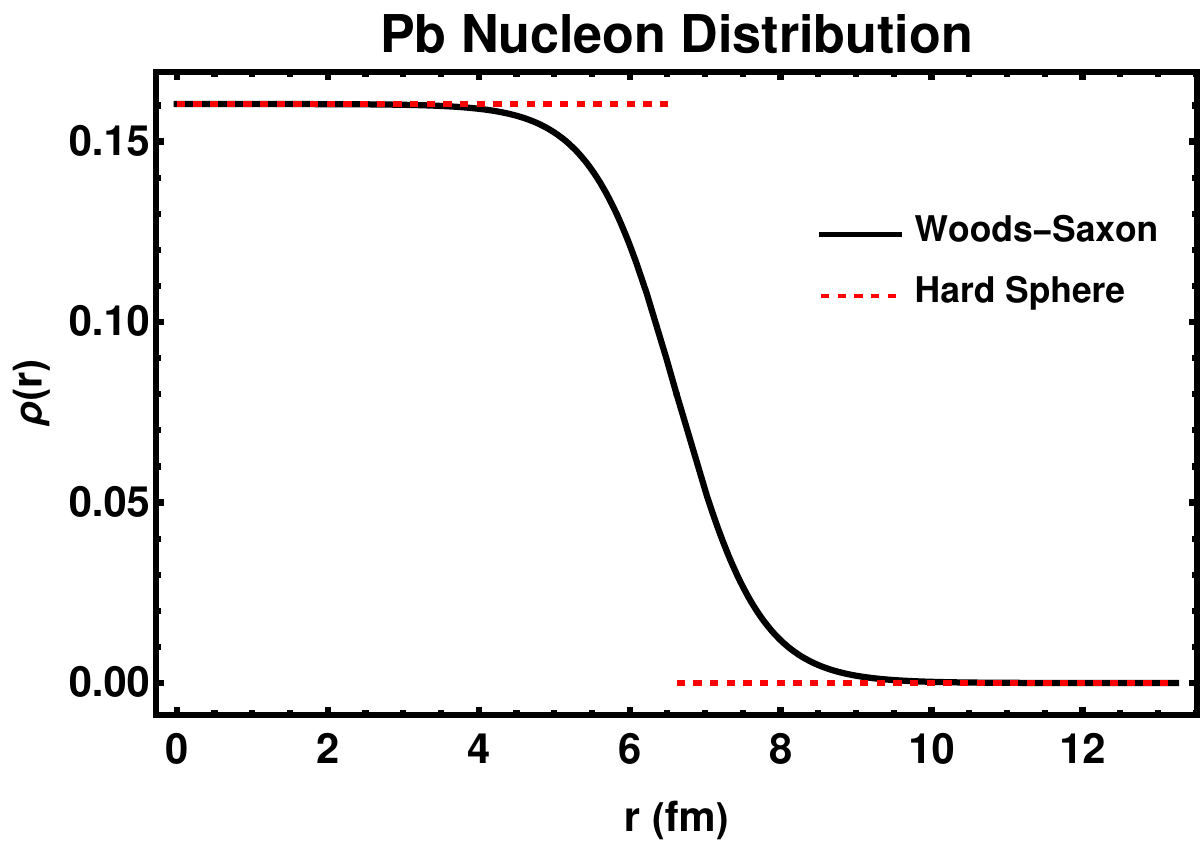}
    \caption{$Pb$ density with $w=0$, $R=6.62 \text{ fm}$ and $a=0.546 \text{ fm}$ \cite{Pbparam}.}
    \label{Pb density}
\end{figure}

The $Pb$ ($A=208$) density distribution is shown in Fig. \ref{Pb density}. The hard-sphere density distribution is also shown in the figure. It can be seen that the nucleon density is larger in the center of the nucleus (for small values of $r$), and decreases as the distance from the center increases.

\begin{figure}[h]
    \centering
    \includegraphics[scale=0.6]{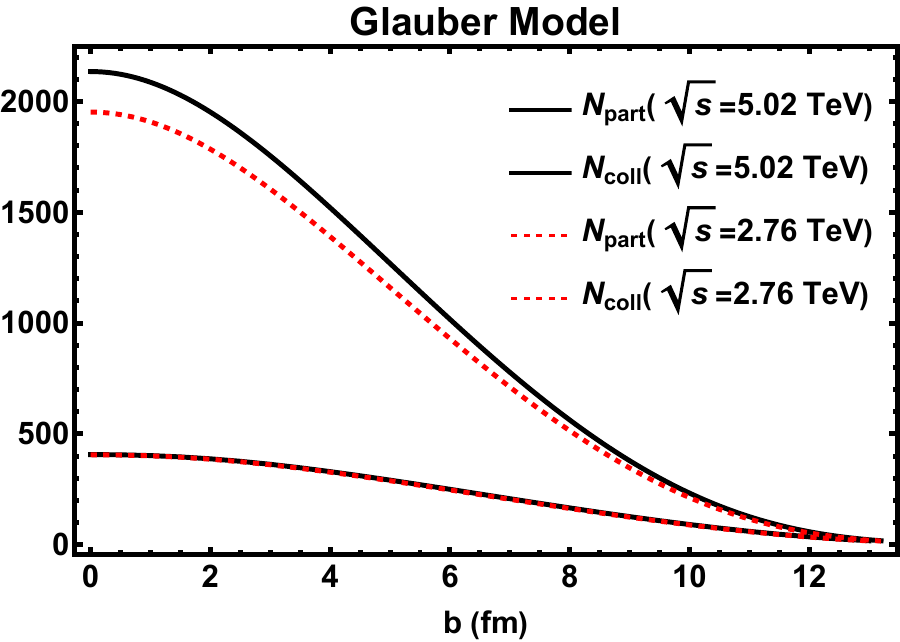}
    \caption{$N_{part}$ and $N_{coll}$  dependence on b for $Pb-Pb$ collisions at $\sqrt{s}_{NN}=2.76 \text{ TeV}$ and $\sqrt{s}_{NN}=5.02 \text{ TeV}$.}
    \label{NpartNcoll}
\end{figure}

Using the Woods-Saxoon distribution in the formalism developed in the previous section for $Pb-Pb$ collisions at $\sqrt{s}_{NN}=2.76 \text{ TeV}$ and $\sqrt{s}_{NN}=5.02 \text{ TeV}$, specially in Eq.(\ref{Ncoll}) and Eq.(\ref{Npart}), the model leads to the results shown in Fig. \ref{NpartNcoll}. The inelastic nucleon cross sections used in the calculations are $\sigma_{inel}^{NN}=64 \text{ mb}$ and $\sigma_{inel}^{NN}=70 \text{ mb}$ \cite{sigmaNNPb}, respectively. As it was expected, the larger values of $N_{part}$ and $N_{coll}$ occur for small $b$ collisions (UC collisions), and decrease as $b$ increases, becoming almost null for $b \approx 2R$ (ultra-peripheral collisions). In central collisions the matter superposition is much bigger than in ultra-peripheral collisions. Also, it can be seen that $N_{part}$, which is approximated to $A+B$ (total number of nucleons) for $b=0$, does not change with $\sqrt{s}_{NN}$, but $N_{coll}$ does.

It would be useful to represent these quantities as functions of the  central rapidity density to adapt the model to the experimental language, and test its applicability. According to \cite{Kharzeev-2001}, the central pseudorapidity density can be written as 

\begin{equation}
    \frac{dN}{d\eta}=n_{pp}(s) \big\{(1-f)\frac{N_{part}(b)}{2} + fN_{coll}(b)\big\}
    \label{dndeta}
\end{equation} where the $f$ parameter is chosen to calibrate the hard particle production, with $(1-f)$ making reference to the soft process fraction; $n_{pp}(s)$ is the multiplicity density, given by \cite{npp}

\begin{equation}
    n_{pp}=2.5-0.25 \log[s] + 0.023 (\log[s])^2
    \label{npp}
\end{equation}

\begin{figure}[h]
    \centering
    \includegraphics[scale=0.6]{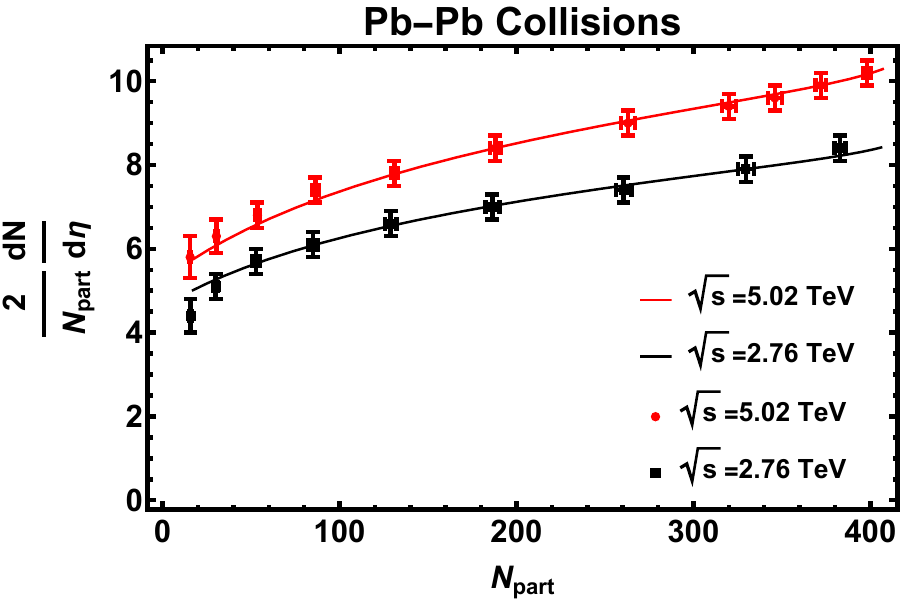}
    \caption{$Pb-Pb$ charged  pseudorapidity density at midrapidity collision data. The experimental data are from \cite{Alice-2.76} and \cite{Alice-5.02}.}
    \label{pbpb 2 e 5}
\end{figure}

For $Pb-Pb$ collisions at  $\sqrt{s}=2.76 \text{ TeV}$ and $ \sqrt{s}=5.02 \text{ TeV}$, the experimental and theoretical data are plotted in Fig. \ref{pbpb 2 e 5}. The central pseudorapidity density is calculated from Eq.(\ref{dndeta}) and $N_{part}$, $N_{coll}$, with the Glauber model shown in Fig. \ref{NpartNcoll}. For the first center of mass energy, the hard fraction was set to $f=10\%$, and for the second, $f=11.5 \%$. The charged particle pseudorapidity density at midrapidity can be well fitted by the Glauber model and Eq.(\ref{dndeta}).

\section{pp Collisions}

The Glauber model developed for $AA$ collisions can be adapted to $pp$ collisions by making some considerations. In the line of the development made in the previous section, the first intuitive approach could be approximating the proton as a nucleus composed by one nucleon. In principle, it is not necessarily wrong, but the Woods-Saxon density profile can hide some important features about the internal structure of the proton, like charge and anisotropic distribution of mass, and subnucleonic effects become more important \cite{loi16,woundedq}. 

\begin{figure}[h!]
\centering
\begin{tabular}{ccc}
    \includegraphics[scale=0.4]{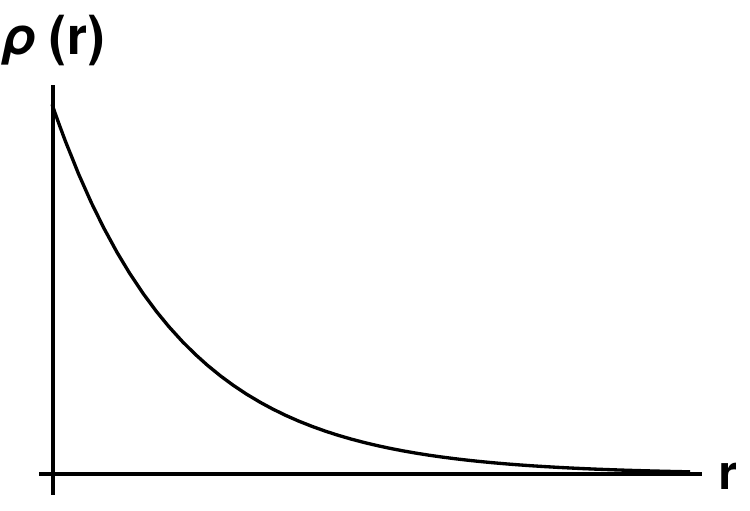}& 
\,\,\,\,\,\,\,\,\,\,\,\,\,\,\,\,\,\,\,\,\,\,\,\, 
\,\,\,\,\,
                                              & 
    \includegraphics[scale=0.4]{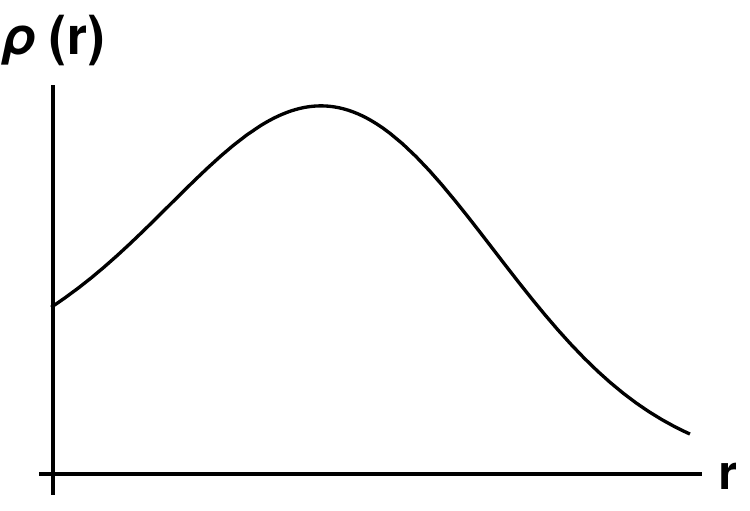} \\
  (a) & \,\,\, & (b)
\end{tabular}
    \caption{a) Charge distribution of the proton measured in elastic 
electron proton scattering at low energies and $Q^2 \simeq 0$. 
b) Charge distribution suggested by the Y shape model of the proton.}
\label{rhocha}
\end{figure}

In \cite{gla11,gla12,gla17} the Renormalization Group Procedure for Effective Particles (REGPEP) was proposed as a way to deal with the spatial evolution of the proton structure for $Q^2\gg \Lambda_{QCD}^2$. In this approach the proton evolves to a configuration with three effective quarks (quarks + antiquarks + gluons), disposed in the vertices of an equilateral triangle, with star-like junction (baryon junction) between them. The charge distribution changes, qualitatively, to the one presented in Fig.\ref{rhocha}b. REGPEP had already been successfully applied to explain other aspects of pp collisions, like anisotropic flow and multiplicity \cite{kubi14,kubi15,gla16}.

Taking all the electromagnetic and gluon effects discussed in chapter 2 into account, then going back to the REGPEP, the subnucleonic approach of Glauber model leads to a different meaning of $N_{part}$ and $N_{coll}$. They change to ``number of partons that participate in the collision'' and ``number of binary collisions between partons'' for a given impact parameter $b$. Also, all the approximations made to the nucleon motion in the nucleus in the Glauber Model are transposed to the parton motion in the proton. That is: because of the high energy of the protons, the partons carry sufficient momentum to be mainly undeflected while the protons pass through each other. Also, their motion is considered to be independent from the proton one and the proton size is much bigger than the range of parton-parton forces.

The parton density can be given by\cite{kubi14}:

\begin{equation} 
\rho_p(\mathbf{r};\mathbf{r}_1,\mathbf{r}_2,\mathbf{r}_3)= \sum_{i=1}^3  
\rho_q(\mathbf{r}-\mathbf{r}_i)+\rho_g \bigg(\mathbf{r}-\sum_{i=1}^3 
\frac{\mathbf{r}_i}{3} \bigg)
	\label{rhop}
\end{equation}
where $i=1,2,3$ are the three effective quarks, with the Gaussian distribution

\begin{equation}
	\rho_q(r)=(1-\kappa)\frac{N_g}{3}
\frac{e^{-r^2/2r_q^2}}{(2\pi)^{3/2}r_q^3}
	\label{rhoq}
\end{equation} 
and a Gaussian gluon distribution, peaked in the average coordinate of $\mathbf{r}_1$, $\mathbf{r}_2$ and $\mathbf{r}_3$, given by:

\begin{equation}
	\rho_g(r)=\kappa N_g \frac{e^{-r^2/2r_g^2}}{(2\pi)^{3/2}r_g^3}
	\label{rhog}
\end{equation} 

\begin{figure}
\begin{tabular}{ccc}
    \includegraphics[scale=0.68]{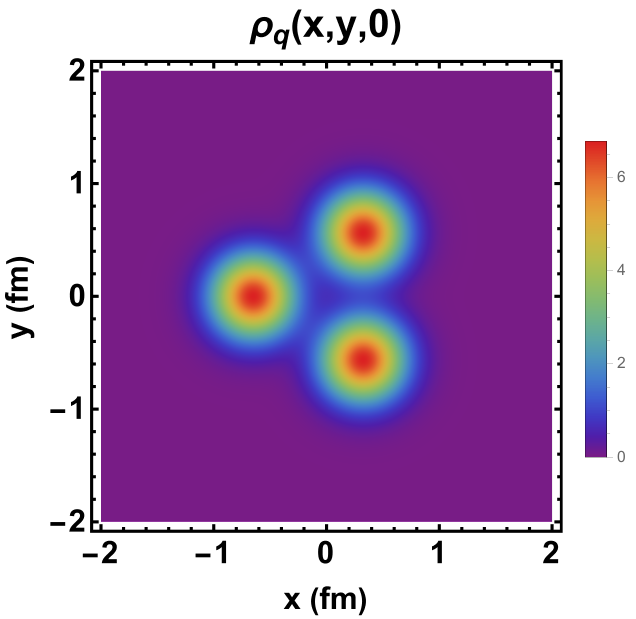}& 

                                              & 
    \includegraphics[scale=0.68]{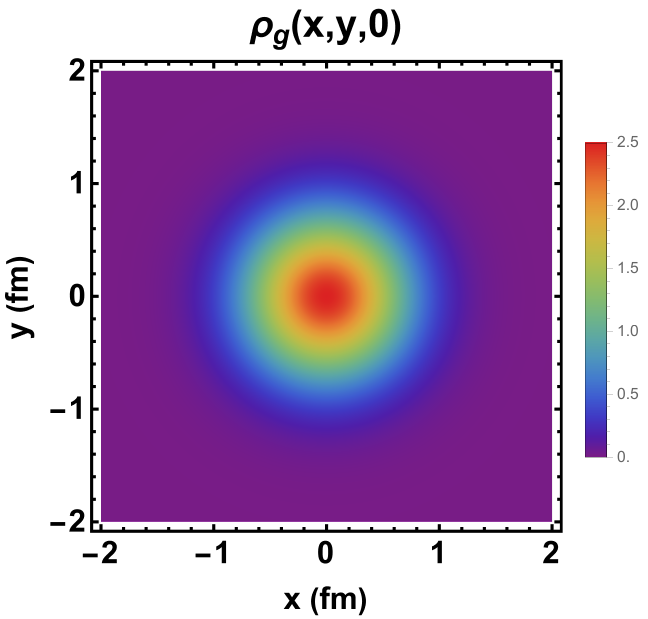} \\
  (a) & \,\,\, & (b)
\end{tabular}
    \caption{a) Quark  and b) gluon  distributions in the transverse plane for the parameters of set 1 on Table \ref{param-glauset}.}
\label{rhos1}
\end{figure}

In the previous formulas, the parameters are: the radius of the effective quarks $r_q$; the extension of the gluon body $r_g$; the degrees of freedom, or number of partons, $N_g$; the fraction of partons that corresponds to gluon $\kappa$. The density plots of the distributions in the transverse plane are shown in Fig.\ref{rhos1}.

\begin{figure}[!ht]
        \centering
\includegraphics[scale=0.8]{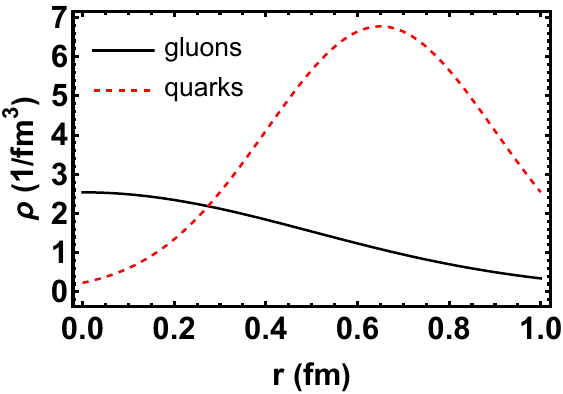}
        \caption{Quark and gluon distributions for the parameters of set 1 on Table \ref{param-glauset}. Projection of Fig.~\ref{rhos1}
along a diagonal direction, starting from the center of the proton and 
passing through one maximum of the quark density.}
        \label{rhos2}
\end{figure}

Fig.\ref{rhos2} illustrates the ``core-corona" aspect of the distribution. In the center (core) the gluon distribution dominates, but in the periphery (corona), the quark distribution is greater. This configuration has some consequences that will be better explored later.

Adapting the Glauber formalism, the probability
per area of finding a parton in the proton flux tube is given by

\begin{equation}
	T_p (x,y)= \int \rho_p (x,y,z) dz =T_p^q (x,y)+ T_p^g (x,y)
	\label{Tp}
\end{equation} 
with $T_p^q (x,y)$ and $T_p^g (x,y)$ from the quark and gluon terms 
of Eq.(\ref{rhop}). Explicitly, they are given by: 
\begin{equation}
T_p^q(x,y)= \frac{N_g(1-\kappa)}{6\pi r_q^2} \sum_{i=1}^3 
e^{-\frac{(x-x_i)^2+(y-y_i)^2}{2r_q^2}}
	\label{Tpq}
\end{equation}
and
\begin{equation}
T_p^g(x,y)=\frac{N_g\kappa}{2\pi r_g^2} e^{-\frac{(x-\sum_{i=1}^3x_i/3)^2
+(y-\sum_{i=1}^3y_i/3)^2}{2rg^2}}
	\label{Tpg}
\end{equation}
The thickness function is given by

\begin{equation}
T_{pp\prime}(b)=\int T_p(x-b/2,y)T_{p\prime}(x+b/2,y)dxdy
\label{Tpp}
\end{equation}
This equation can be splitted into four analytic terms. The interaction between the quarks $i=1,2,3$ from $p$ and quarks $j=1,2,3$ from $p\prime$ (the quark-quark term), is given by: 

\begin{eqnarray}
T_{pp\prime}^{qq}(b) &=& \int T_p^q(x-b/2,y) T_{p\prime}^q(x+b/2,y) dxdy
 \nonumber \\
                     &=& \frac{N_g^2 (1-\kappa)}{36\pi r_q^2}          
\sum_{i=1}^3\sum_{j=1}^3 \exp \bigg(-\frac{\big(b+(x_i-x_j)\big)^2
+\big(y_i-y_j\big)^2}{4 rq^2} \bigg)
\label{Tqq}
\end{eqnarray}
The gluon-gluon term can be calculated as:

\begin{eqnarray}
&&T_{pp\prime}^{gg}(b) = \int T_p^g(x-b/2,y) T_{p\prime}^g(x+b/2,y) dxdy 
  \\
                     &=& \frac{N_g^2 \kappa^2 }{4\pi r_g^2} \exp     
\bigg(-\frac{\big(b+\sum_{i=1}^3 x_i/3 -\sum_{j=1}^3 x_j/3\big)^2 
+\big(\sum_{i=1}^3 y_i/3 -\sum_{j=1}^3 y_j/3\big)^2}{4rg^2}\bigg)\nonumber
\label{Tgg}
\end{eqnarray}
The gluon-quark term (interactions between the gluons from $p$ and quarks from $p\prime$) is calculated as:

\begin{eqnarray}
T_{pp\prime}^{gq}(b) &=& \int T_p^g(x-b/2,y) T_{p\prime}^q(x+b/2,y) dxdy
 \\
&=& \frac{ N_g^2 \kappa (1-\kappa)}{6\pi (r_q^2+r_g^2)} 
\sum_{j=1}^3 \exp \bigg(-\frac{\big(b+\sum_{i=1}^3 x_i/3 -x_j \big)^2 
+ \big(\sum_{i=1}^3 y_i/3-y_j\big)^2}{2(r_q^2+r_g^2)}\bigg)\nonumber
\label{Tgq}
\end{eqnarray}
Finally, the quark-gluon term is:

\begin{eqnarray}
T_{pp\prime}^{qg}(b) &=& \int T_p^q(x-b/2,y) T_{p\prime}^g(x+b/2,y) dxdy
 \\
                     &=& \frac{N_g^2 \kappa (1-\kappa)}{6\pi (r_q^2+r_g^2)}  
\sum_{i=1}^3 \exp \bigg(-\frac{\big(b+x_i-\sum_{j=1}^3 x_j/3   \big)^2 + 
\big(y_i-\sum_{j=1}^3 y_j/3\big)^2}{2(r_q^2+r_g^2)}\bigg)\nonumber 
\label{Tqg}
\end{eqnarray}
In the end, inserting the terms above into Eq.(\ref{Tpp}), the total thickness is: 

\begin{equation}
T_{pp\prime}(b)=T_{pp\prime}^{qq}(b)+T_{pp\prime}^{gg}(b)+T_{pp\prime}^{gq}(b)
+ T_{pp\prime}^{qg}(b)
\label{Tppfinal}
\end{equation}
The number of binary collisions between partons is: 

\begin{equation}
N_{coll}=T_{pp\prime}(b)\sigma^{pp\prime}
\label{Ncollp}
\end{equation} 
where $\sigma^{pp\prime}$ is the parton-parton cross section and the number of participants is calculated according to Eq.(\ref{Npart estimado}).

\begin{figure}
    \centering
    \includegraphics[scale=0.3]{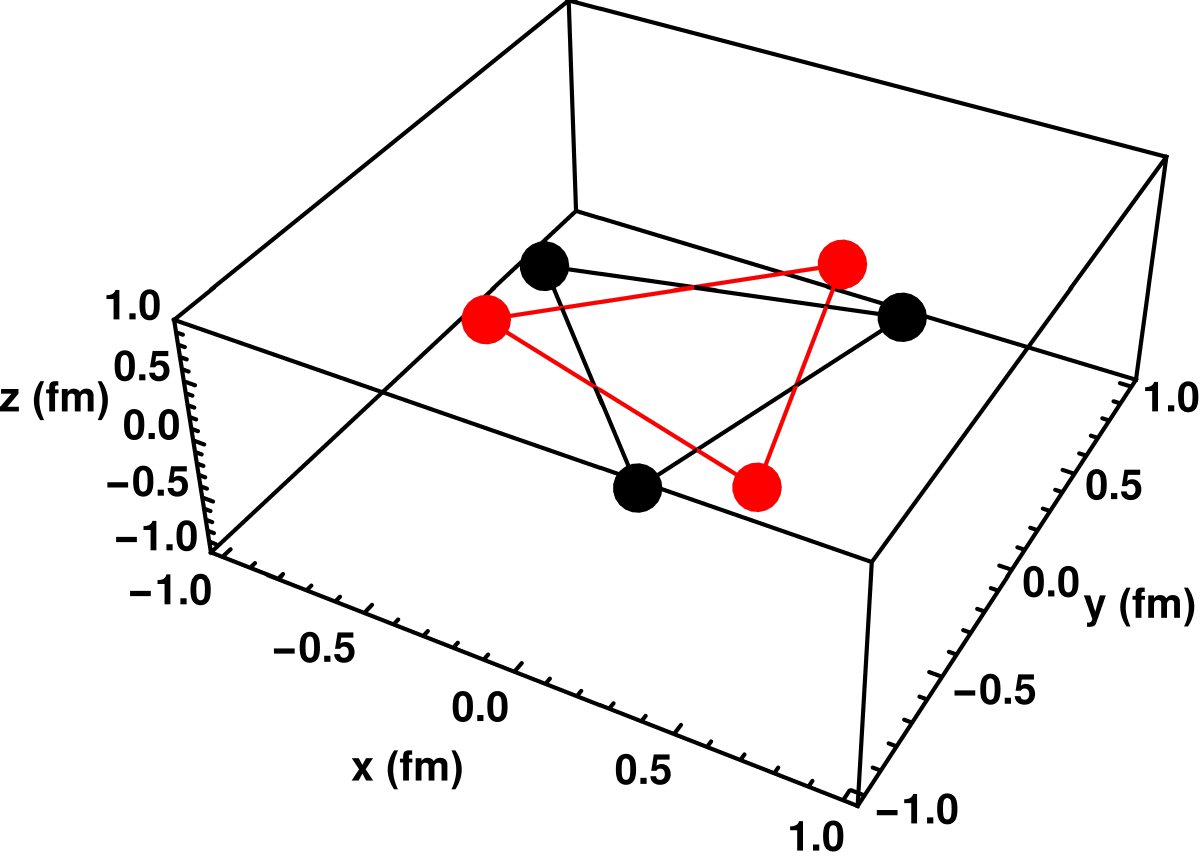}
    \caption{Example of initial spatial configuration of the two 
colliding protons.}
    \label{initialcond}
\end{figure}

Since the calculation above needs effective quarks positions inputs in the transverse plane, we will adopt the following procedure: the quarks $\mathbf{r}_i$ from the projectile $p$, and $\mathbf{r}\prime_i$ from the target $p\prime$, will be positioned according to

\begin{equation}
\mathbf{r}_i=\frac{d}{2}\big(\cos(\phi_i + \alpha), 
\sin(\phi_i + \alpha) \big)
\,\,\,\,\,\,\,\,\,\,\,\,\,
\mbox{and}
\,\,\,\,\,\,\,\,\,\,\,\,\,
\mathbf{r'}_i=\frac{d}{2}\big(\cos(\phi_i + \beta), 
\sin(\phi_i + \beta)\big)
\label{ris}
\end{equation}
In the formulas above, $\phi_1 = \pi/3$, $\phi_2 = -\pi/3$ and $\phi_3 = - \pi$. This choice of coordinates positions the quarks in the vertices of an equilateral triangle, with the angles $\alpha$ and $\beta$ randomly chosen in the range $\alpha\text{, }\beta \in  [0,2\pi]$ in order to rotate them around the $z$ axis. In Fig. \ref{initialcond} we show one example of initial condition.

The computational procedure is the following: for a given impact parameter $b$, the angles $\alpha$ and $\beta$ are chosen. After these choices, the coordinates of Eq.(\ref{ris}) are set (initial condition) and thickness function of Eq.(\ref{Tppfinal}) is calculated. Then the thickness is applied in Eq.(\ref{Ncollp}) and, indirectly, in Eq.(\ref{Npart estimado}) to calculate $N_{coll}$ and $N_{part}$. For each impact parameter there are 10000 initial condition choices and the final value for the outputs are averages of $N_{coll}$ and $N_{part}$ of these configurations. The impact parameter is such that $b \in  [0,2.2] \text{ fm}$, with steps 
$\Delta b=0.1 \text{ fm}$.  

\begin{figure}
\centering
\begin{tabular}{c}
    \includegraphics[scale=0.5]{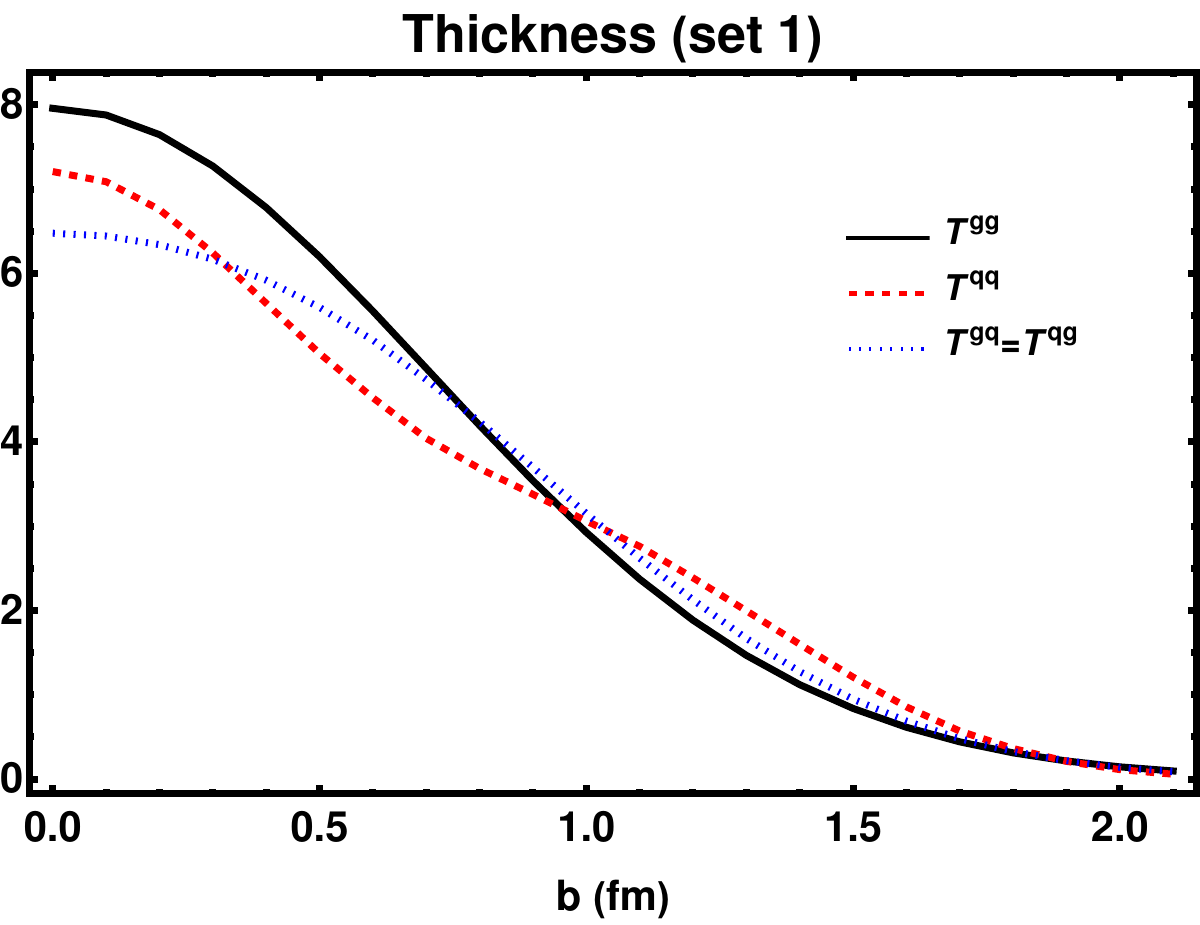}\\ 
    \includegraphics[scale=0.5]{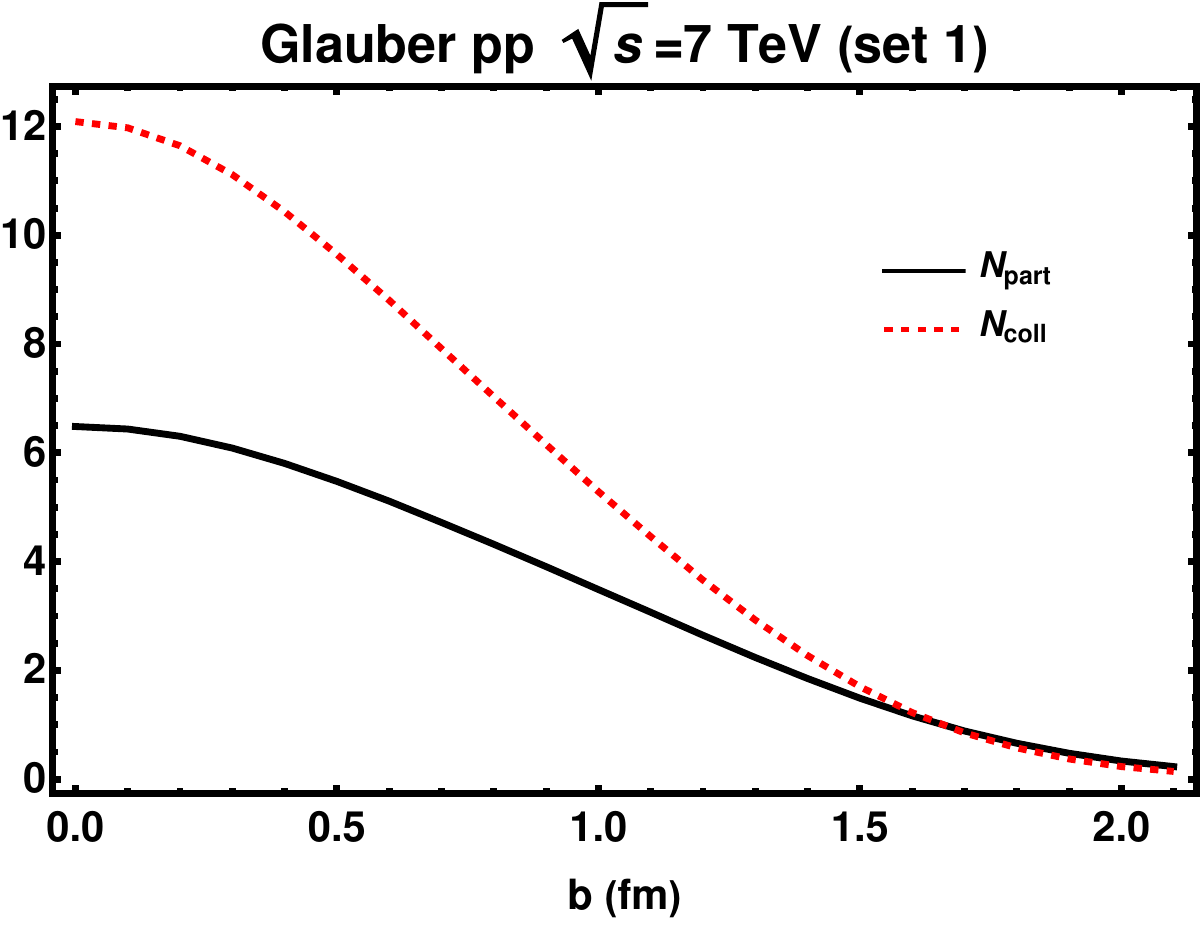} \\
    \includegraphics[scale=0.5]{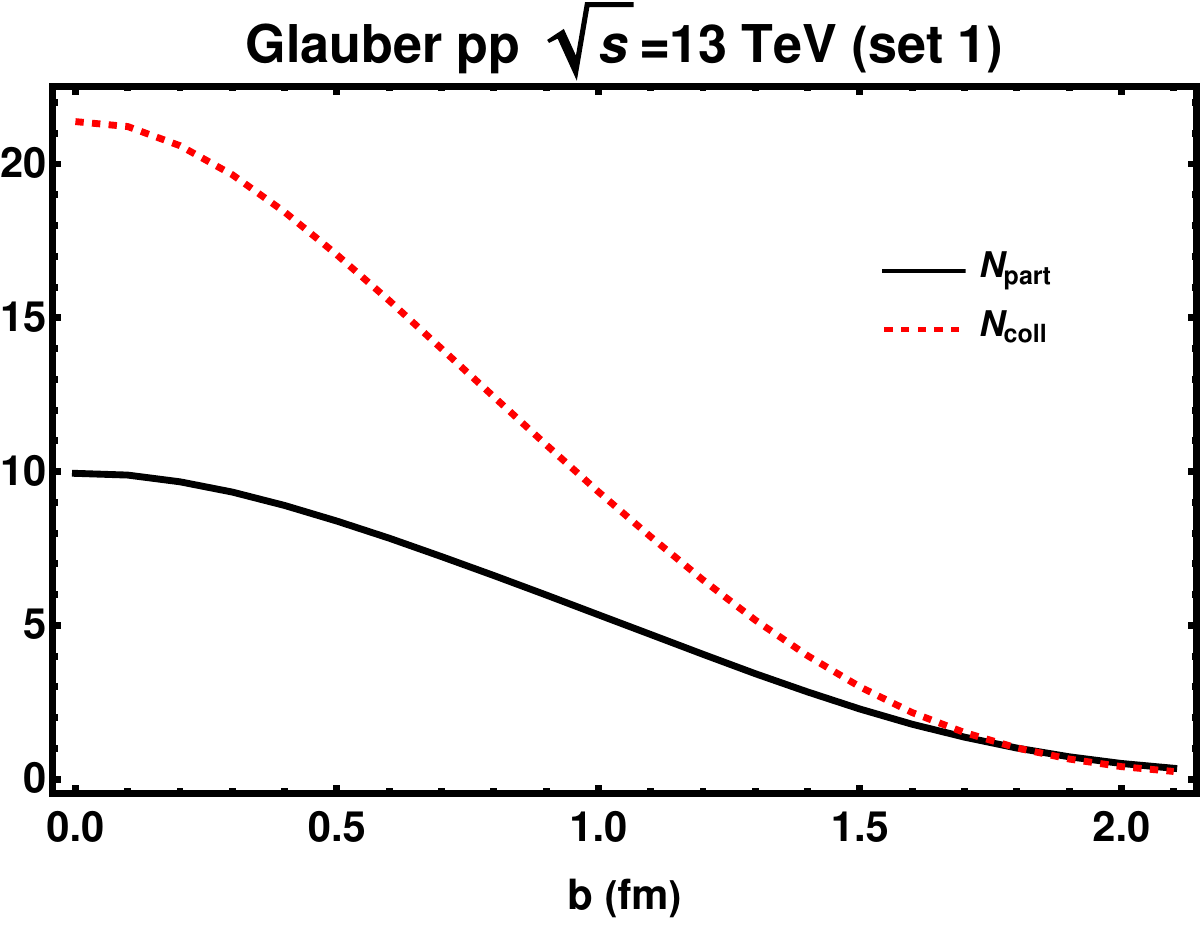}\\

\end{tabular}
    \caption{Thickness functions and average  of $N_{part}$ and $N_{coll}$ over different quark spatial configurations, at $\sqrt{s}=7\text{ TeV}$ and $\sqrt{s}=13\text{ TeV}$, for the parameter set 1.}
\label{ppset1}
\end{figure}

\begin{figure}
\centering
\begin{tabular}{c}
    \includegraphics[scale=0.5]{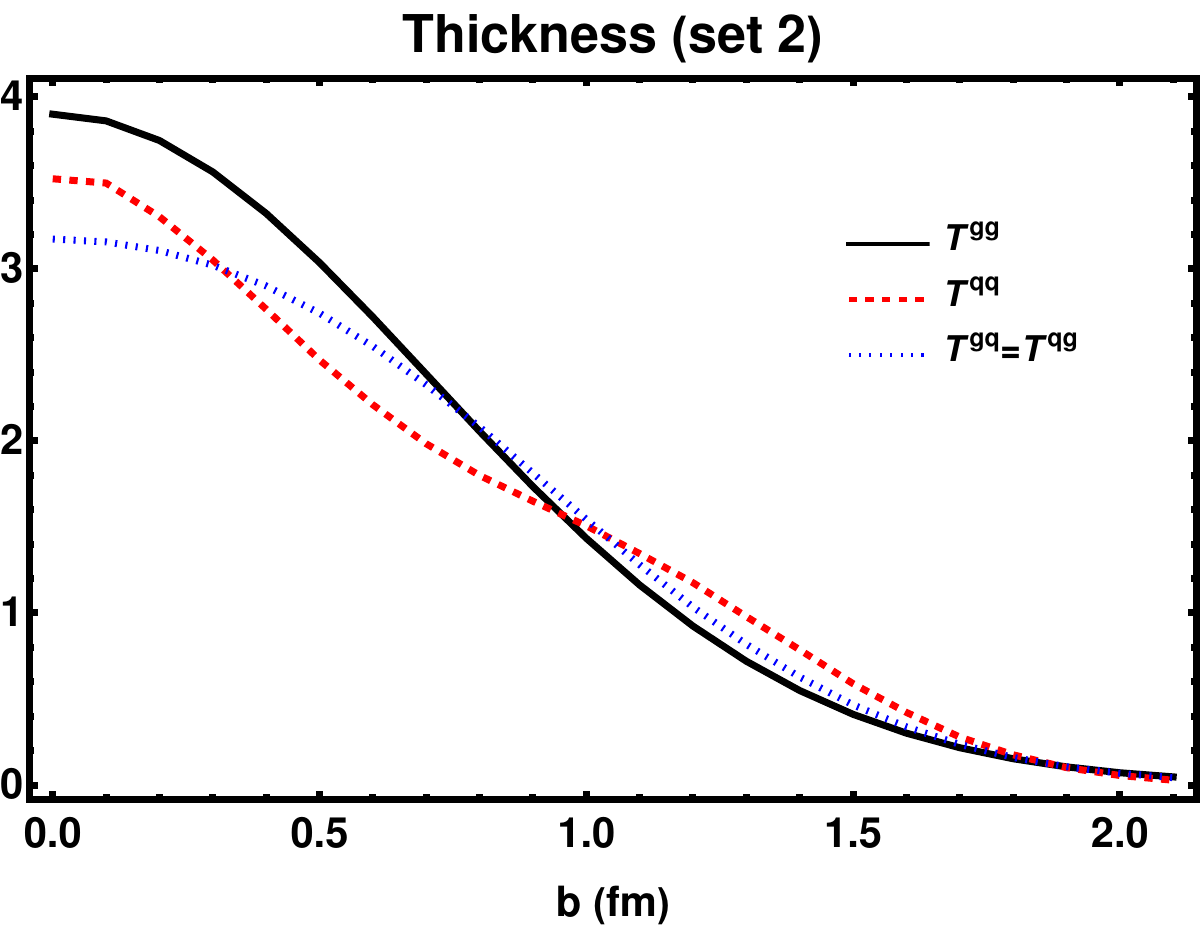}\\ 
    \includegraphics[scale=0.5]{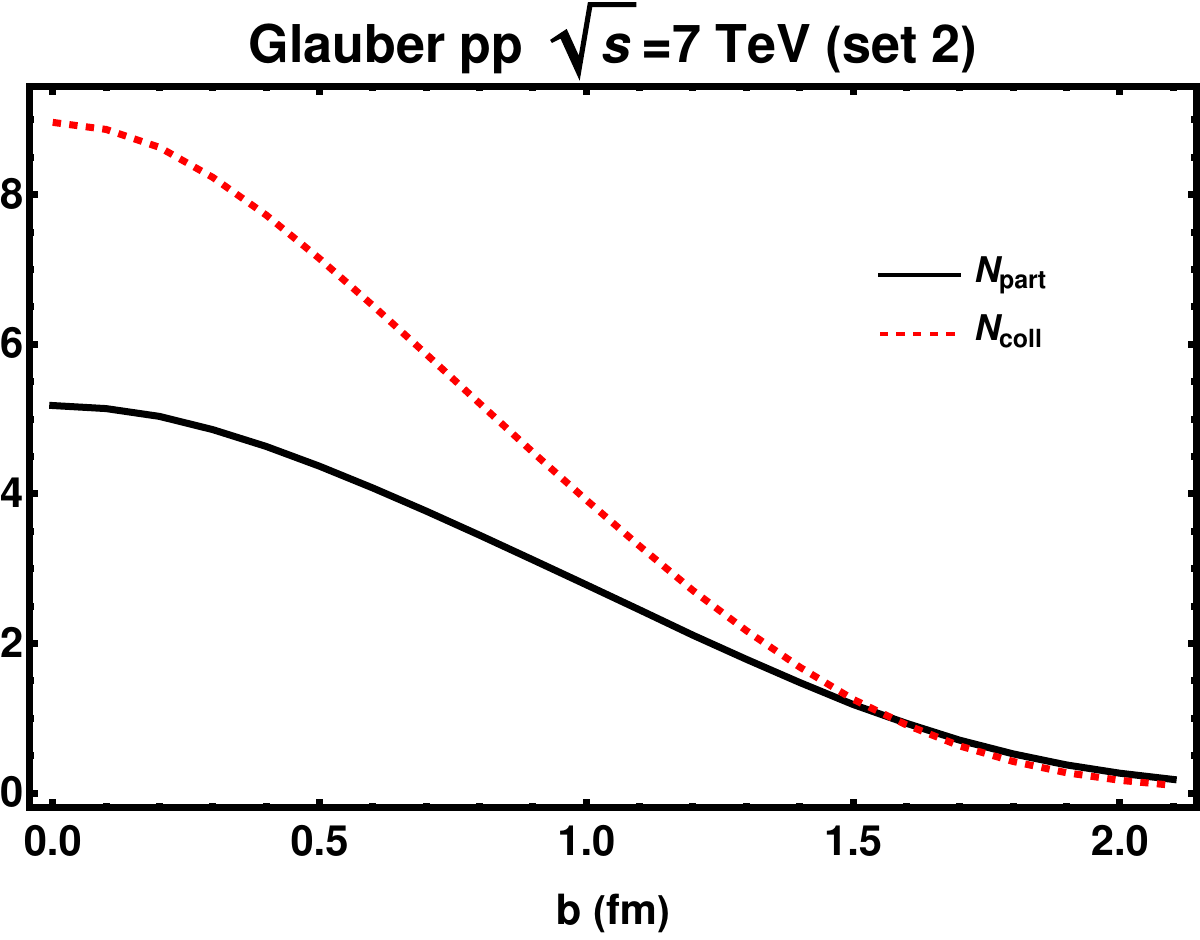} \\
    \includegraphics[scale=0.5]{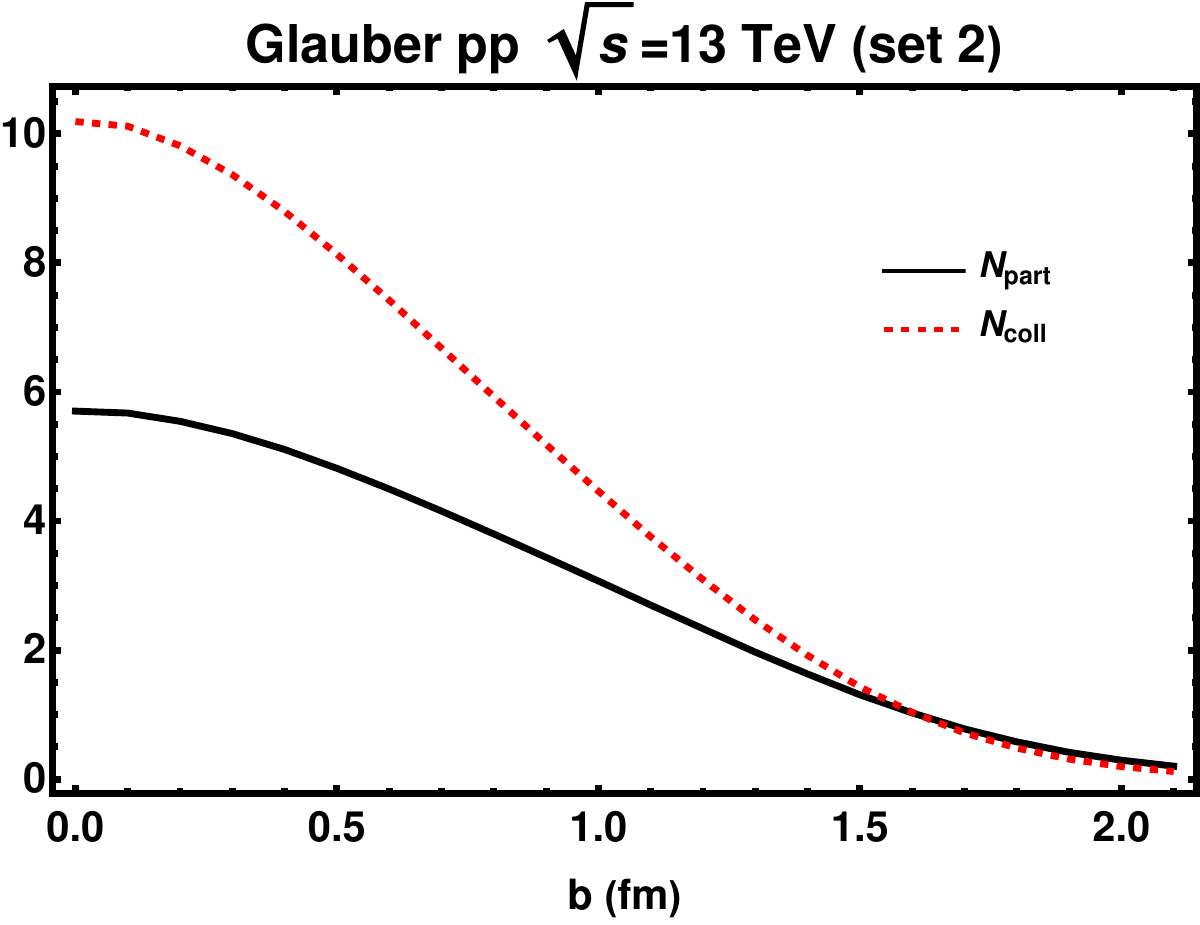}\\

\end{tabular}
    \caption{Thickness functions and average  of $N_{part}$ and $N_{coll}$ over different quark spatial configurations, at $\sqrt{s}=7\text{ TeV}$ and $\sqrt{s}=13\text{ TeV}$, for the parameter set 2.}
\label{ppset2}
\end{figure}

\begin{figure}
\centering
\begin{tabular}{c}
    \includegraphics[scale=0.5]{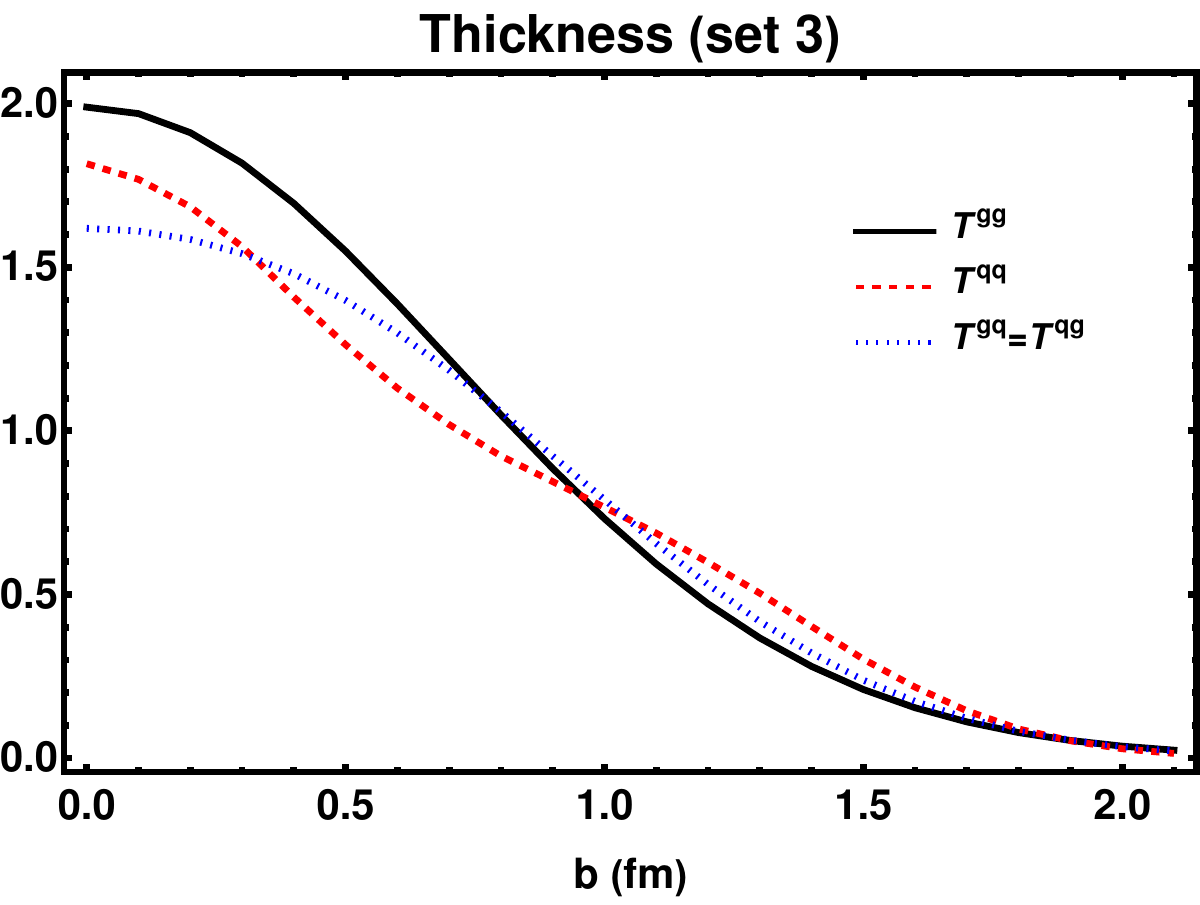}\\ 
    \includegraphics[scale=0.5]{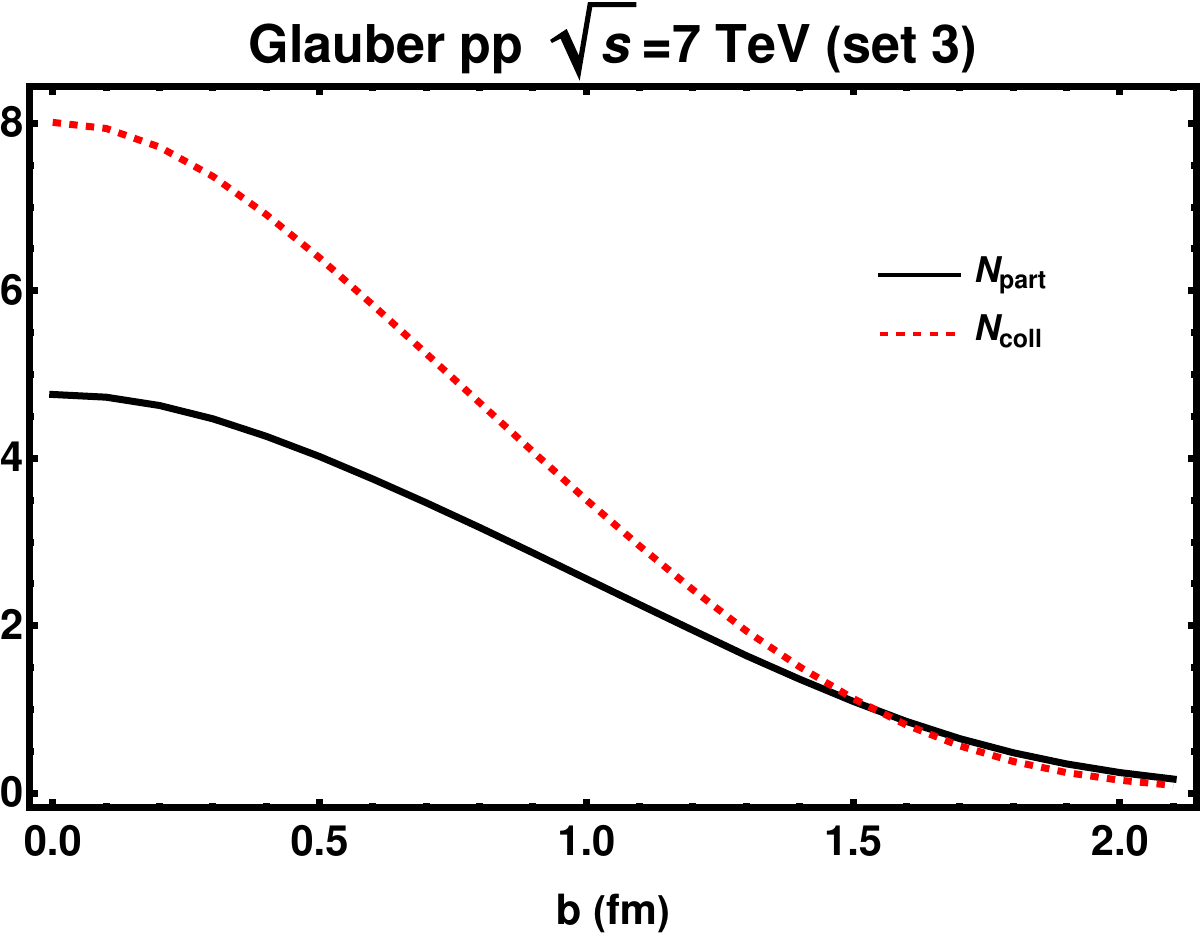} \\
    \includegraphics[scale=0.5]{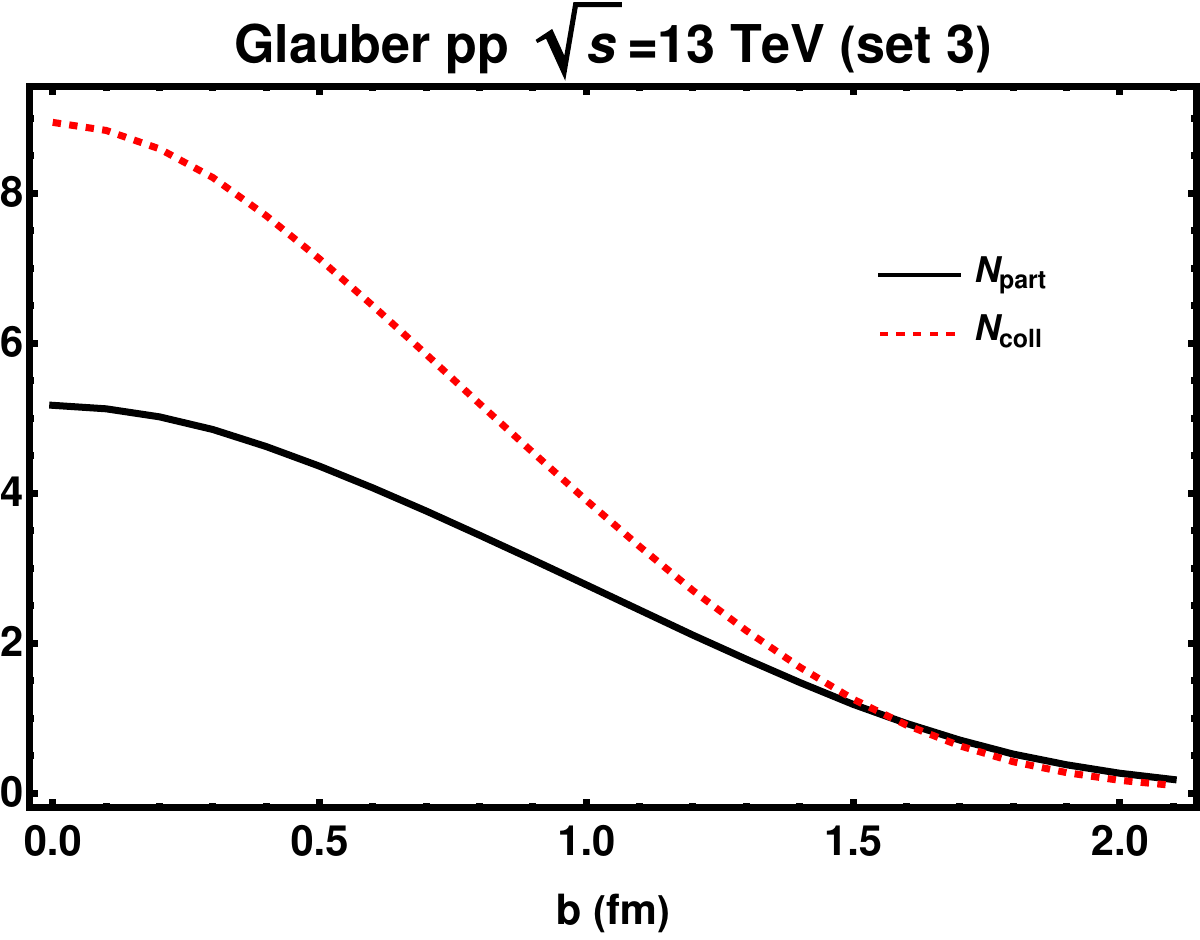}\\

\end{tabular}
    \caption{Thickness functions and average  of $N_{part}$ and $N_{coll}$ over different quark spatial configurations, at $\sqrt{s}=7\text{ TeV}$ and $\sqrt{s}=13\text{ TeV}$, for the parameter set 3.}
\label{ppset3}
\end{figure}

The thickness functions and Glauber outputs for the parameters sets of Table \ref{param-glauset} are shown from Fig. \ref{ppset1} to  \ref{ppset3}. The parameters that do no change from one set to another are $\kappa=0.5$, $r_q=0.25 \text{ fm}$, $d=1.3 \text{ fm}$ \cite{deb20} and $r_g=0.5 \text{ fm}$ \cite{kubi14}. All the parameters were taken from previous works.  These Glauber outputs will be used in charm production that will be better discussed in chapter 5.

\begin{table}[ht!]
\caption{Effective number of partons and parton-parton cross sections used in the Glauber model.}
\label{param-glauset}
\centering
    \begin{tabular}{| c | c | c | c |}
    \hline
     & Set 1 & Set 2 &Set 3\\ \hline
    $N_g$           & $10$ \cite{deb20} & $7$ \cite{loi16} & $5$ \cite{loi16} \\ \hline
    $\sigma^{pp}\text{ (mb)}$ $(5.02\text{ TeV})$  &  $2.8$ \cite{loi16} &  $5.7$ \cite{loi16}&  $10.3$\cite{loi16}   \\ \hline
    $\sigma^{pp}\text{ (mb)}$ $(7\text{ TeV})$  &  $4.3$ \cite{sigmaparton} &  $6.5$ \cite{loi16}&  $11.4$\cite{loi16}   \\ \hline
    $\sigma^{pp}\text{ (mb)}$ $(13\text{ TeV})$ &  $7.6$ \cite{sigmaparton} & $7.4$ \cite{loi16}& $12.7$ \cite{loi16}  \\ \hline
    \end{tabular}
\end{table}

\pagebreak

\section{pA Collisions}

Another interesting situation where the Glauber model can be applied is in proton-nucleus collisions. As in the proton case, subnucleon degrees of freedom become important during the collision, that is, the internal structure of the nucleons can also play an important role. We will attempt to develop the model of the previous sections and adapt it to $p-Pb$ collisions.

\begin{figure}[h!]
    \centering
    \includegraphics[scale=0.7]{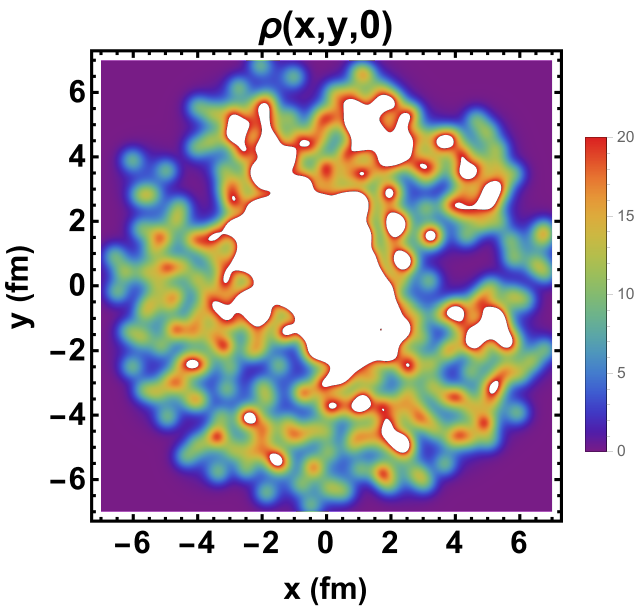}
    \caption{Example of initial spatial configuration in the transverse plane of the $Pb$ nucleus. The white regions are high multiplicity points.}
    \label{Pbinitialcond}
\end{figure}

Assuming that all the nucleons have the same number density distribution of Fig. \ref{rhos1} (Eq.(\ref{rhop})), the positions of the nucleons inside the nucleus, in the transverse plane, can be chosen as follows: the radial position of the nucleon in the transverse plane $s$ and the azimuthal angle $\gamma$ are randomly chosen in the range $0<s<R=6.62 \text{ fm}$ and $0<\gamma <2\pi$. After choosing the radial position, the nucleon can be rotated around the $z$ axis by a angle $\alpha$, such that $0<\alpha<2\pi$, like in the $pp$ case shown in Fig. \ref{initialcond}. The procedure is repeated for the $208$ nucleons. These assumptions, together with the usual Glauber ones, leads to initial conditions like in Fig. \ref{Pbinitialcond}.

Then the $Pb$ density will be calculated as the sum of $208$ densities of Eq.(\ref{rhop}) as:

\begin{equation} 
	\rho_{Pb}(\mathbf{r})= \sum_{i=1}^{208} \rho_{p_i}(\mathbf{r};\mathbf{r}_1,\mathbf{r}_2,\mathbf{r}_3) 
	\label{rhoPb}
\end{equation}
and each quark of each nucleon will be positioned according to

\begin{equation}
	\mathbf{r}_i=\bigg(\frac{d}{2}\cos(\phi_i + \alpha)+s\cos(\gamma)\mathbf{,} \sin(\phi_i + \alpha) +s\sin(\gamma)\bigg)
\end{equation}

Then, proceeding with Glauber Formalism, the probability per area of finding a parton in the flux tube is given by:

\begin{eqnarray}
T_{Pb}(x,y) &=& \int \sum_i^{208} \rho_{p_i}(x,y,z) dz =\sum_i^{208}\big(T_{p_i}^q(x,y)+T_{p_i}^g(x,y) \big)
 \nonumber \\
                     &=& T_{Pb}^q(x,y)+T_{Pb}^g(x,y)
\end{eqnarray}
where $T_{p_i}^q(x,y)$ and $T_{p_i}^g(x,y)$ are given by Eq.(\ref{Tpq}) and Eq.(\ref{Tpg}).

The thickness function is calculated as:

\begin{equation}
	T_{pPb}(b)=\int T_p(x-b/2,y)T_{Pb}(x+b/2,y)dxdy
\label{TpPb}
\end{equation}

As before, the thickness can be splitted into four analytic terms. The quark-quark is:

\begin{eqnarray}
T_{pPb}^{qq}(b) &=& \int T_p^q(x-b/2,y) T_{pPb}^q(x+b/2,y) dxdy
 \nonumber \\
                     &=& \sum_i^{208} T_{pp_i}^{qq}(b)
\label{TqqpPb}
\end{eqnarray}
where $T_{pp_i}^{qq}(b)$ is given by Eq.(\ref{Tqq}). The gluon-gluon term is:

\begin{eqnarray}
T_{pPb}^{gg}(b) &=& \int T_p^g(x-b/2,y) T_{Pb}^g(x+b/2,y) dxdy 
 \nonumber \\
                     &=& \sum_i^{208} T_{pp_i}^{gg}(b)
\label{TggpPb}
\end{eqnarray}
with $T_{pp_i}^{gg}(b)$ given by Eq.(\ref{Tgg}). The gluon-quark and quark-gluon terms are, respectively:

\begin{eqnarray}
T_{pPb}^{gq}(b) &=& \int T_p^g(x-b/2,y) T_{Pb}^q(x+b/2,y) dxdy
\nonumber \\
&=& \sum_i^{208} T_{pp_i}^{gq}(b)
\label{TgqpPb}
\end{eqnarray}
and

\begin{eqnarray}
T_{pPb}^{qg}(b) &=& \int T_p^q(x-b/2,y) T_{Pb}^g(x+b/2,y) dxdy
\nonumber  \\
                     &=& \sum_i^{208} T_{pp_i}^{qg}(b)
\label{TqgpPb}
\end{eqnarray}
where  $T_{pp_i}^{gq}(b)$ and $T_{pp_i}^{qg}(b)$ are given by Eq.(\ref{Tgq}) and Eq.(\ref{Tqg}).

The thickness is then calculated as:

\begin{eqnarray}
T_{pPb}(b) &=& T_{pPb}^{qq}(b)+T_{pPb}^{gg}(b)+T_{pPb}^{gq}(b)+T_{pPb}^{qg}(b)
\nonumber  \\
                     &=& \sum_i^{208} \big(T_{pp_i}^{qq}(b)+T_{pp_i}^{gg}(b)+T_{pp_i}^{gq}(b)+T_{pp_i}^{qg}(b) \big)
\label{ThicknespPb}
\end{eqnarray}
In the end, the thickness can be written as individual interactions between the projectile proton and the target nucleons that form the heavy nucleus. Also like before, the number of binary collisions can be calculated as 

\begin{equation}
N_{coll}(b)=T_{pPb}(b)\sigma^{pp\prime}
\label{NcollpPb}
\end{equation} 
where $\sigma^{pp\prime}$ is the parton-parton inelastic cross section and $N_{part}(b)$ can be approximated as Eq.(\ref{Npart estimado}).

\begin{figure}
\centering
\begin{tabular}{c}
    \includegraphics[scale=0.7]{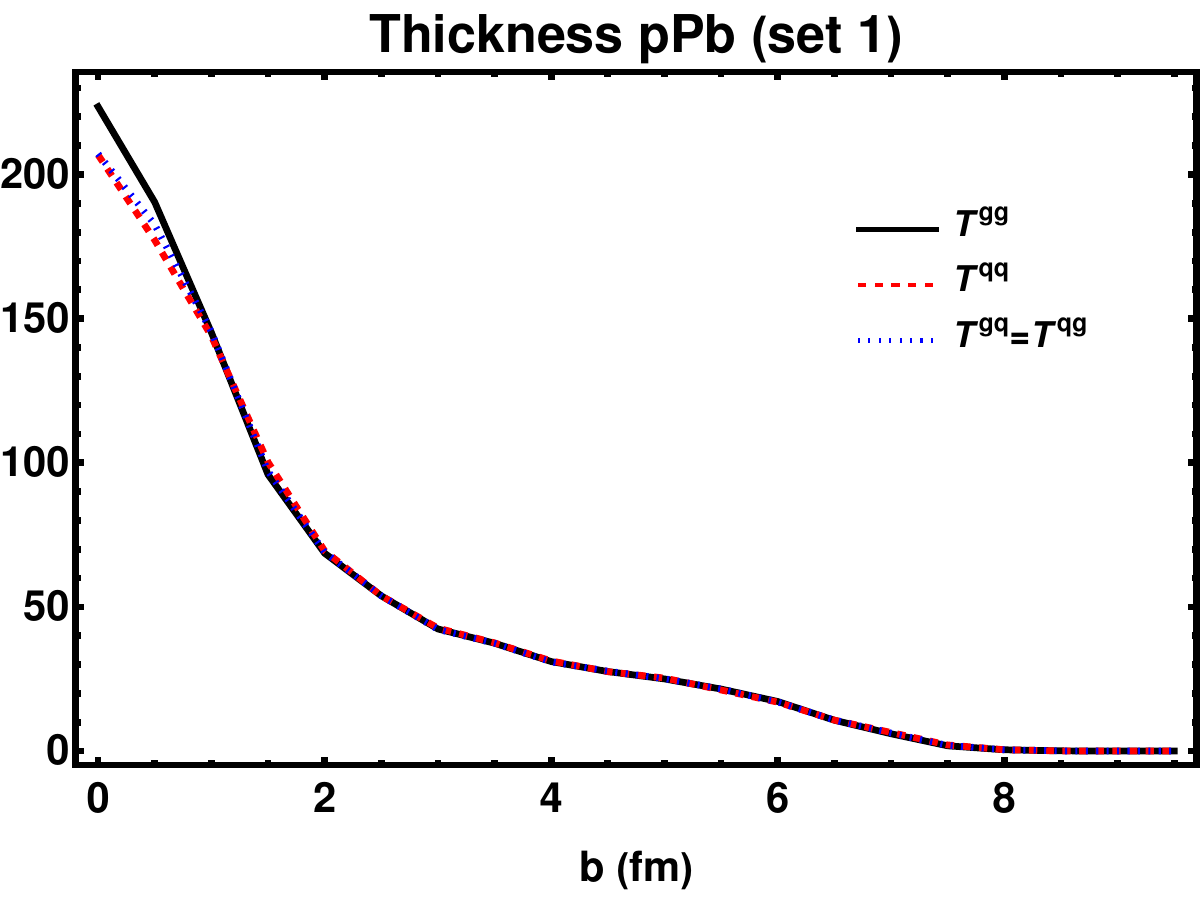}\\ 
    \includegraphics[scale=0.7]{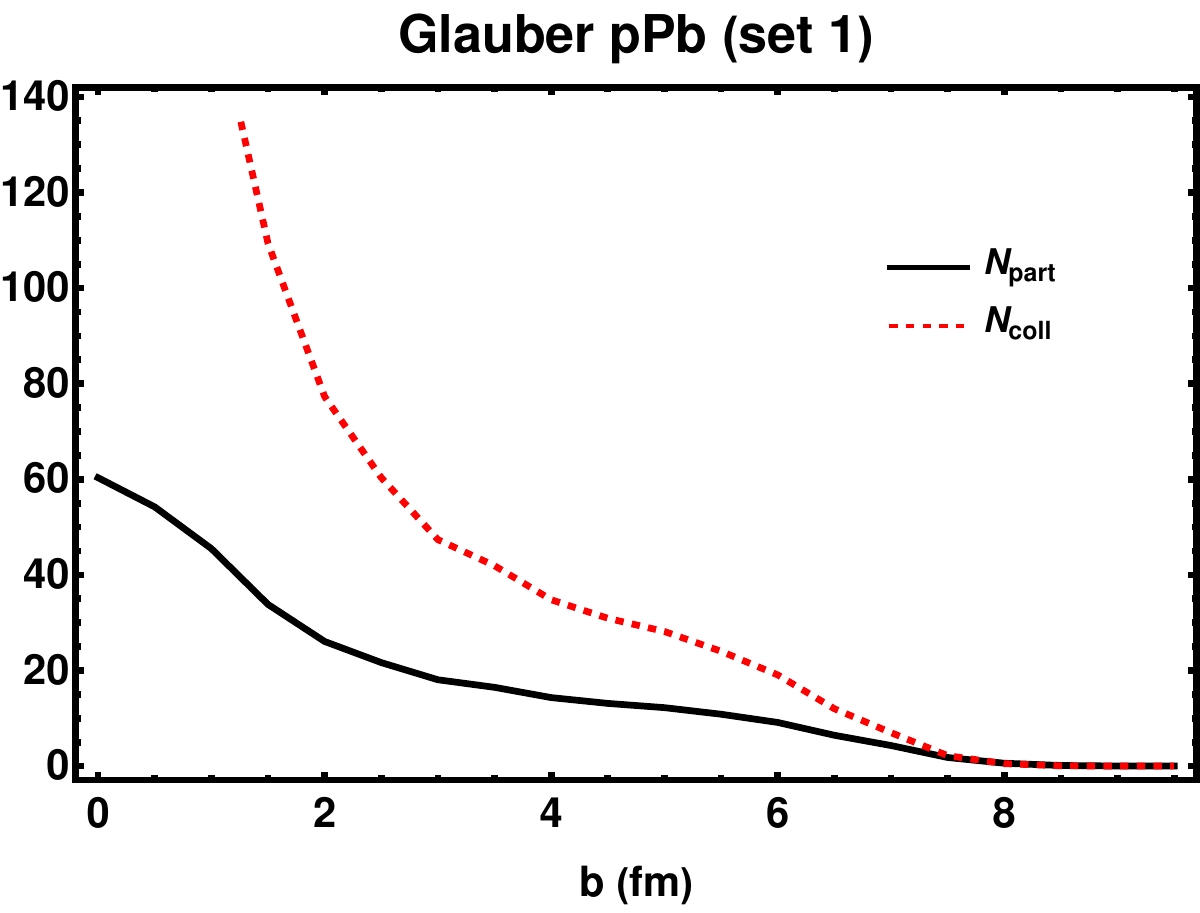} \\

\end{tabular}
    \caption{Thickness functions and average  of $N_{part}$ and $N_{coll}$ over different quark and nucleon spatial configurations, at $\sqrt{s}=5.02\text{ TeV}$, for the parameter set 1.}
\label{pPbset1}
\end{figure}

\begin{figure}
\centering
\begin{tabular}{c}
    \includegraphics[scale=0.7]{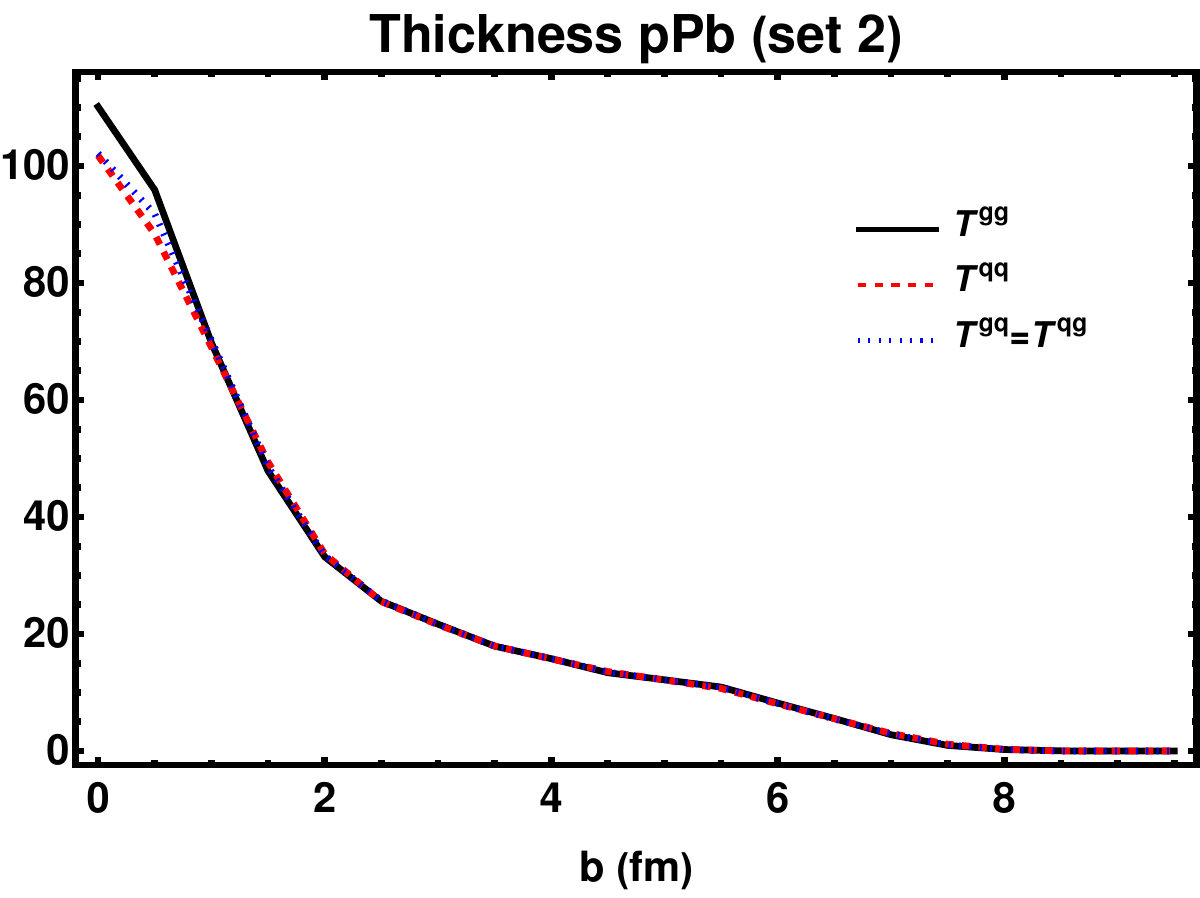}\\ 
    \includegraphics[scale=0.7]{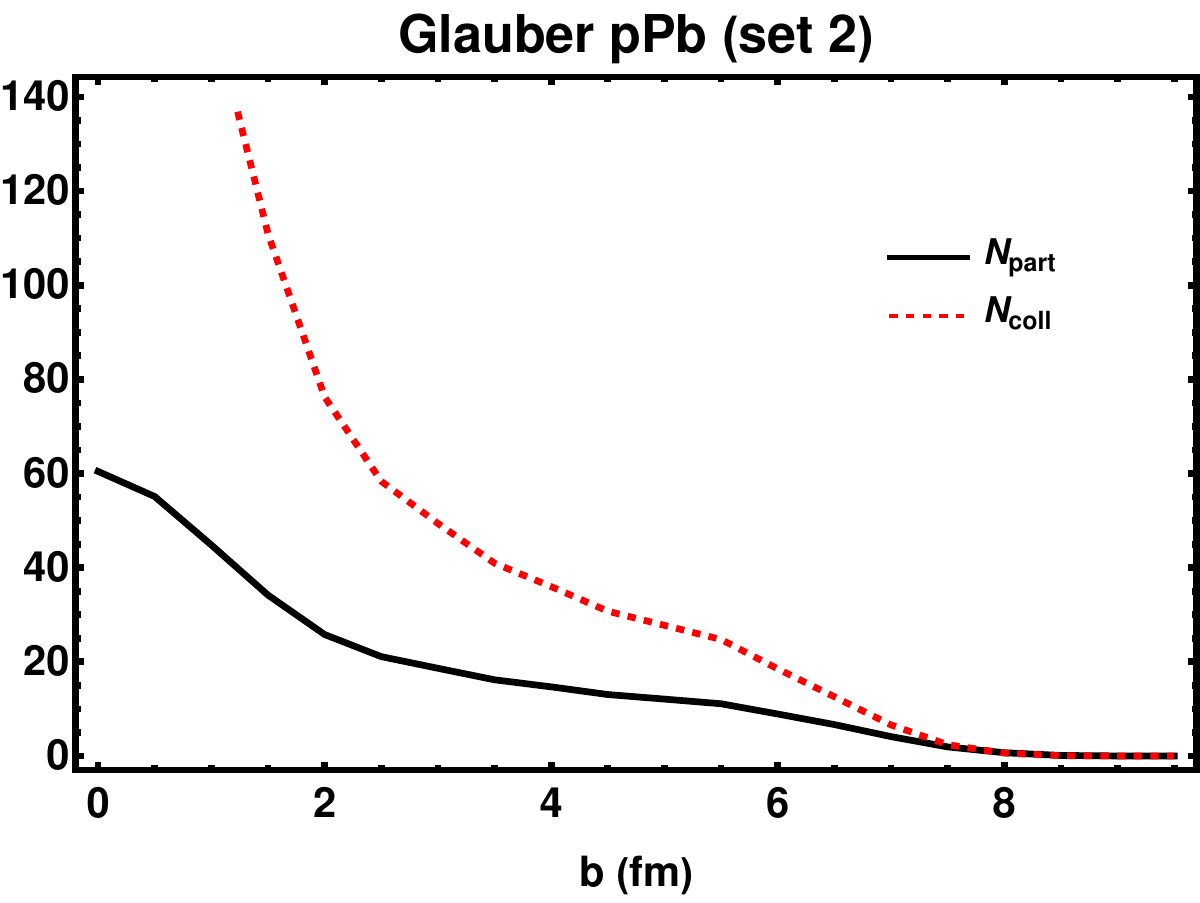} \\

\end{tabular}
    \caption{Thickness functions and average  of $N_{part}$ and $N_{coll}$ over different quark and nucleon spatial configurations, at $\sqrt{s}=5.02\text{ TeV}$, for the parameter set 2.}
\label{pPbset2}
\end{figure}

\begin{figure}
\centering
\begin{tabular}{c}
    \includegraphics[scale=0.7]{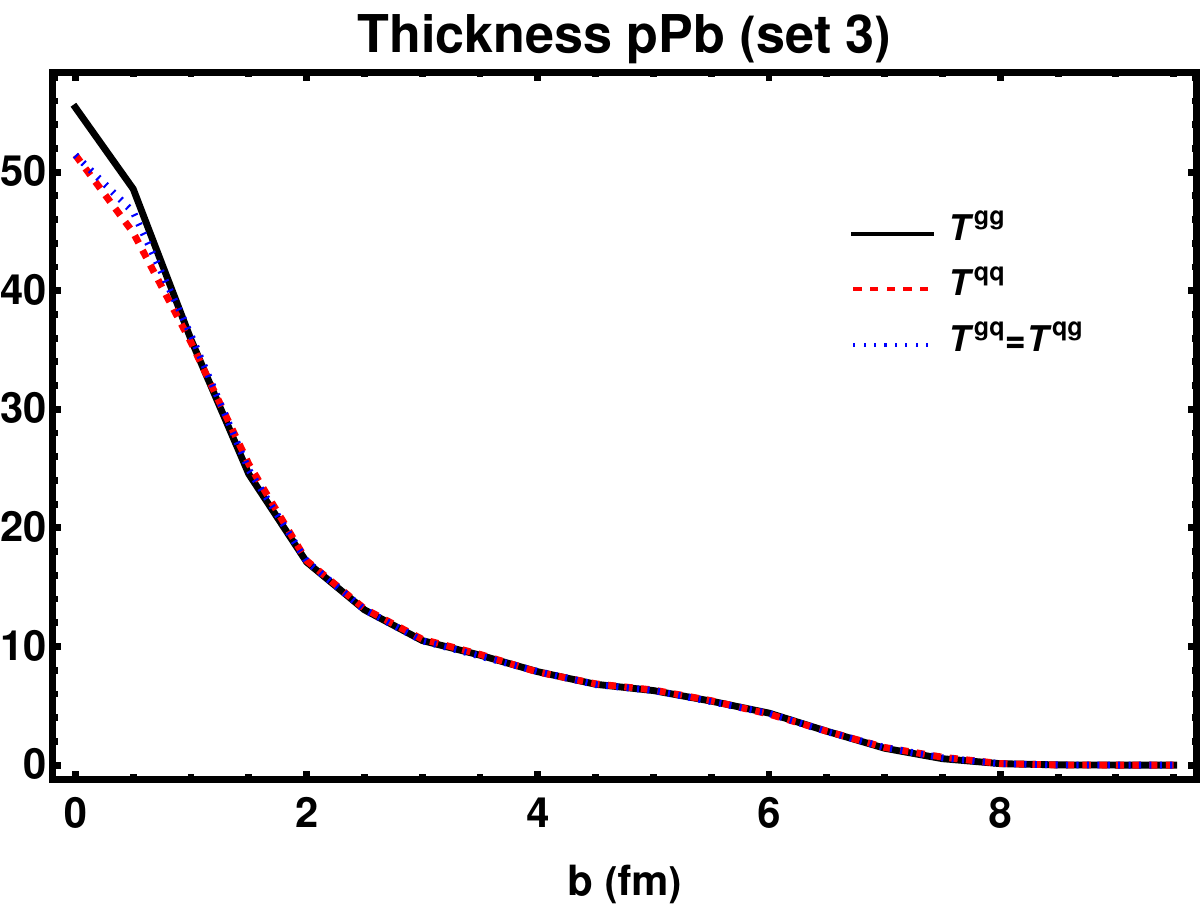}\\ 
    \includegraphics[scale=0.7]{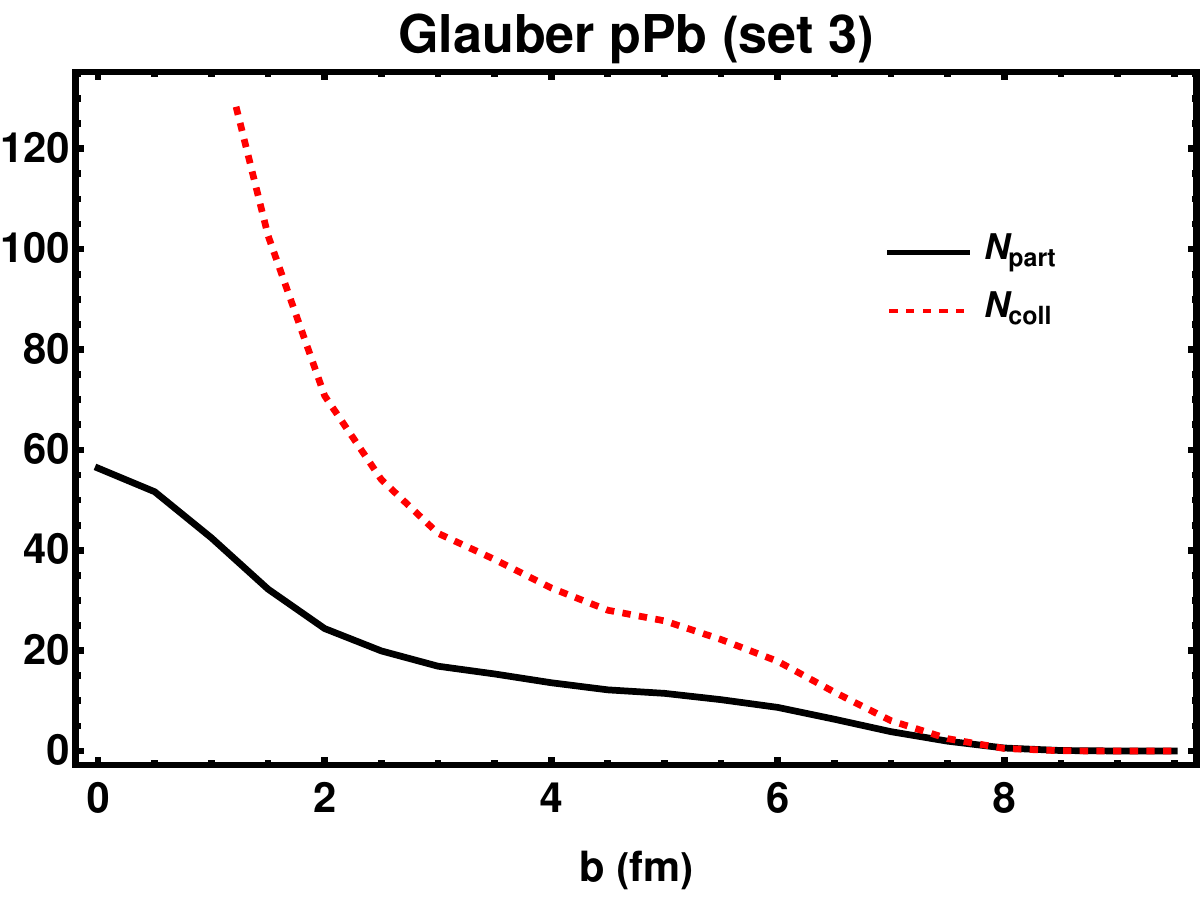} \\

\end{tabular}
    \caption{Thickness functions and average  of $N_{part}$ and $N_{coll}$ over different quark and nucleon spatial configurations, at $\sqrt{s}=5.02\text{ TeV}$, for the parameter set 3.}
\label{pPbset3}
\end{figure}

The thickness and Glauber outputs for $pPb$ collisions for the parameter sets of Table \ref{param-glauset}, averaged over $200$ initial conditions for each impact parameter, are shown from Fig. \ref{pPbset1} to \ref{pPbset3}. The computational proceeding is almost the same adopted in the $pp$ case, but with the choices of $s$ and $\gamma$ for each nucleon inside the nucleus and the impact parameter is in the range $[0,10]\text{ fm}$, with steps $\Delta b=0.5 \text{ fm}$.

\pagebreak

\chapter{Charm Production}

In this work, the charmonium production will by accounted for by the Color Evaporation Model (CEM) \cite{vogt,vogt2}. In this model, the mass of the produced heavy quark pair must be between twice the quark mass and twice the lightest meson that can be formed with the heavy quark, and since the pair production has a non-perturbative component, one assumes that color neutrality is achieved by the ``evaporation" of initial colour fields. In the end, the quarkonium cross section is written in terms  of the partonic cross section and the overlap function. Therefore  the model works as a good tool to study the influence of the internal structure of the colliding hadrons in heavy quark production.

Understanding  quarkonium production is still a work in progress and there is no precise description valid for all kinematic ranges. For example, CEM overshoots data in the large transverse-mass regime \cite{puzzle,puzle2}. The calculation presented below will be integrated in $p_T$ and we expect that the high $p_T$ discrepancy will not strongly affect the results because the curve falls rapidly in this regime.

\section{Charm production in the CEM}

Considering that $c\bar{c}$ pairs are produced through the process $gg \rightarrow c\bar{c}$  and $q \bar{q} \rightarrow c\bar{c}$, the total cross section can be written as:

\begin{eqnarray}
\sigma^{CEM} &=& \mathbf{F} K \sum_{i,j}\int_{(2m_c)^2}^{(\Lambda)^2} dm^2 
\int dx_1 dx_2 f_i(x_1,\mu_F^2) f_j(x_2,\mu_F^2) \sigma_{ij}(m^2,\mu_R^2)  
\delta(m^2-x_1 x_2 s)
\nonumber  \\
             &=& \sigma_{gg}^{CEM} + \sigma_{q\bar{q}}^{CEM}
\label{sigcem}
\end{eqnarray} 
In this total cross section, $f_i(x_1,\mu_F^2)$ and $f_j(x_2,\mu_F^2)$ are parton distribution functions calculated at $\mu_F$ (factorization scale), $\sigma_{ij}(m^2,\mu_R^2)$ are the parton-parton cross sections calculated at $\mu_R$ (renormalization scale) and $F$ is the percentage of $c\bar{c}$ which becomes $J/\psi$. The integration limit $\Lambda$ works as a cut-off. For open charm production, $\Lambda=\sqrt{s}$, and for $J/\psi$, $\Lambda=2m_D=2(1.8)\text{ GeV}^2$.

For our purposes the leading order elementary cross sections will be sufficient. For the relevant processes, they are given by

\begin{equation}
\sigma_{gg}(m^2,\mu_R^2)=\frac{\pi \alpha_s^2(\mu_R^2)}{3m^2} \bigg\{  
\big(1+\frac{4m_c^2}{m^2}+\frac{m_c^4}{m^4} \big) \ln\bigg( 
\frac{1+\lambda}{1-\lambda}\bigg) -\frac{1}{4} \big(7+
\frac{31 m_c^2}{m^2}\big) \lambda \bigg\}
\label{sigmagg}
\end{equation}
and 
\begin{equation}
\sigma_{q\bar{q}}(\mu_R^2)=\frac{8\pi \alpha_s^2(m^2)}{27m^2} 
\bigg(1+\frac{2m_c^2}{m^2}\bigg)\lambda
\label{sigmaqq}
\end{equation}
The parameter $K$ is introduced to include the effect of higher order corrections. The coupling constant is given  by 

\begin{equation}
    \alpha_s(\mu_R^2)=\frac{12 \pi}{(33-2n_f)\ln \big( \frac{\mu_R^2}{\Lambda^2}\big)}
    \label{alpha}
\end{equation} where $n_f$ is the number of flavours involved in the collision and, tipically,  $\Lambda=200 \text{ MeV}$. The renormalization scale will be chosen to be equal to the factorization one, which will be twice the mass of charm quark ($\mu_R=\mu_F=\mu=2m_c)$. The $\lambda$ term is defined as

\begin{equation}
    \lambda =\bigg( 1 - \frac{(2m_c)^2}{m^2} \bigg)^{1/2}
    \label{lambda cem}
\end{equation} 
Imposing that $\lambda$ is real and applying the $\delta$ function properties in Eq.(\ref{sigcem}), the integration limits must satisfy

\begin{equation}
    1-\frac{4m_c^2}{x_1 x_2 s} \ge 0
\end{equation} Noticing that the smallest value of $x_1$ occurs for $x_2=1$, the parton momentum fractions  must be such that $(2m_c)^2/s \le x_1 \le 1$ and $(2m_c)^2/(sx_1) \le x_2 \le 1$. In the end, the multiplicity is given by 

\begin{equation}
    N_{CEM}^{c\bar{c}}=T_{AB}(b) \sigma^{CEM}
    \label{N cem}
\end{equation}

\section{Open Charm Production}

\begin{figure}[h]
    \centering
    \includegraphics[scale=0.55]{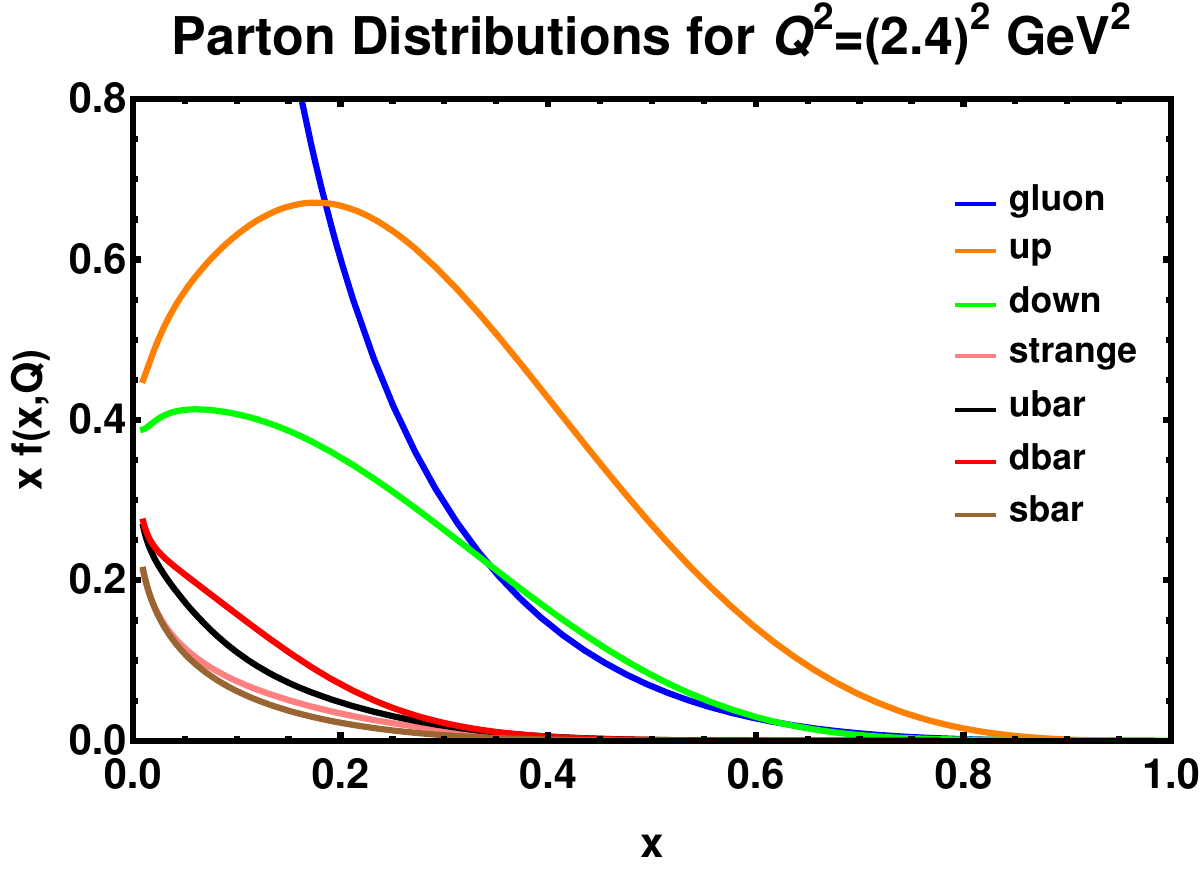}
    \caption{PDFs for $Q^2=(2.4)^2 \text{ GeV}^2$ \cite{PDFs}.}
    \label{PDFs CEM}
\end{figure}

The open charm production cross section can be calculated by taking out the parameter $F$ in Eq.(\ref{sigcem}) with the parton distribution functions from the MMHT 14  set \cite{PDFs}. For $Q^2=\mu^2=(2m_c)^2$, the PDFs are plotted in Fig. \ref{PDFs CEM}. 

\begin{figure}[h!]
	\centering
    \includegraphics[scale=0.6]{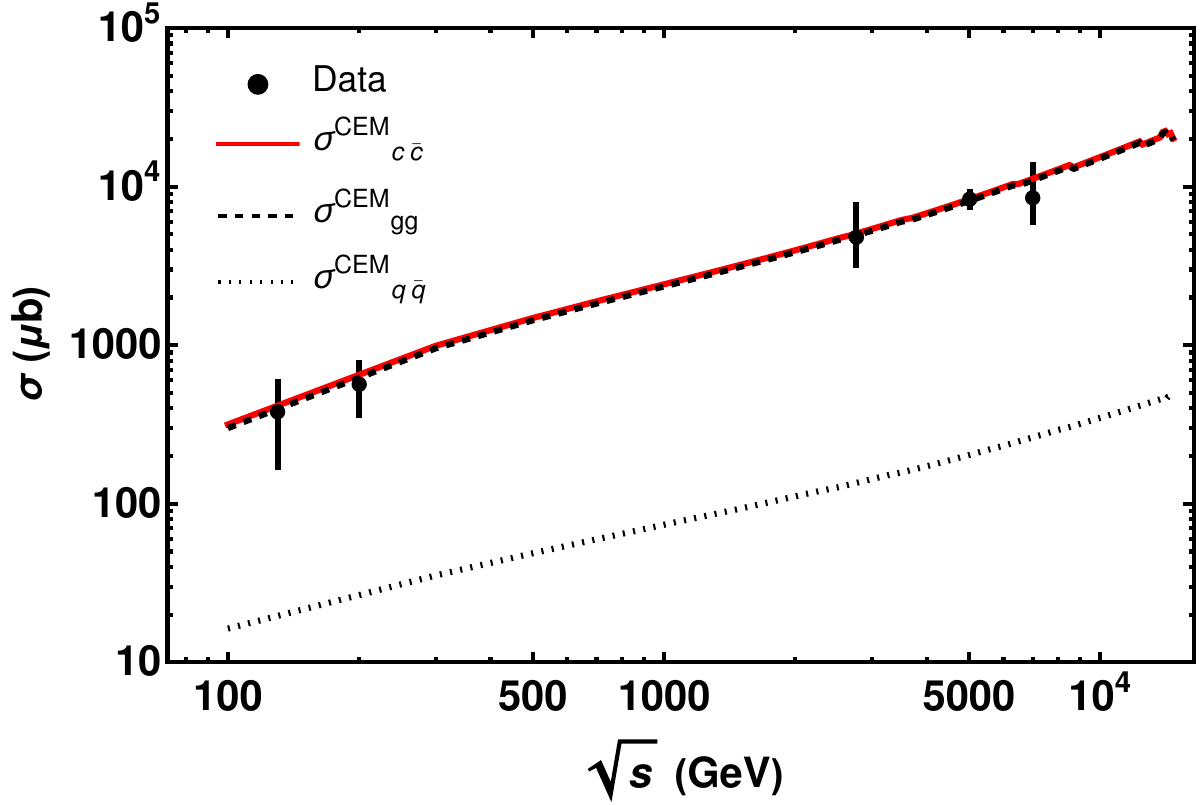}
    \caption{Charm production cross sections for $m_c=1.2 \text{ GeV}$. Contribution of $gg \rightarrow c\bar{c}$ (dashed line), $q \bar{q} \rightarrow c\bar{c}$ (dotted line) and total cross section (solid line). The experimental data are from \cite{phenix02,phenix06,alice12,alicepp5.02}.}
\label{crosscc1}
\end{figure}

\begin{figure}[h!]
	\centering
    \includegraphics[scale=0.6]{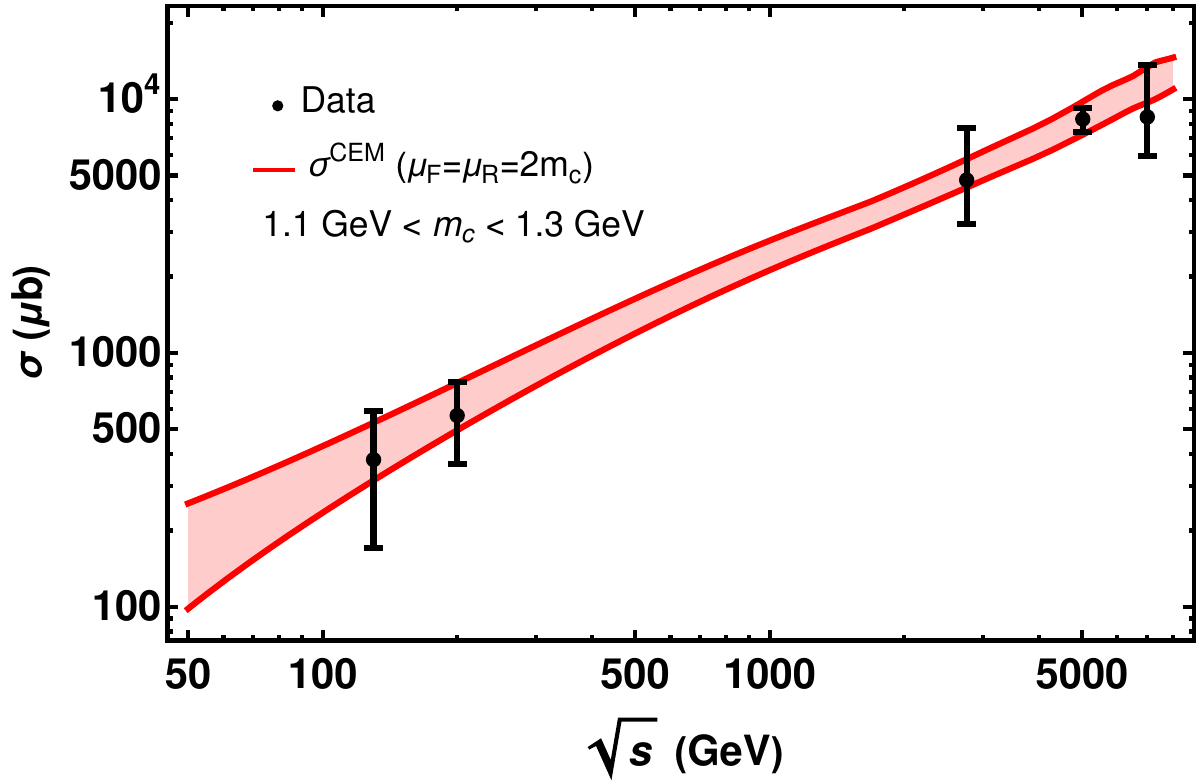}
    \caption{Charm production cross sections for 
$1.1\text{ GeV}<m_c<1.3 \text{ GeV}$. The experimental data are from 
\cite{phenix02,phenix06,alice12,alicepp5.02}.}
\label{crosscc2}
\end{figure}

In order to fix the $K$ parameter of Eq.(\ref{sigcem}), the open charm $c\bar{c}$ cross section is fitted to existing data, as shown in  to Fig.\ref{crosscc1}. The optimal value is $K=3$. As can be seen in the figure, $\sigma_{gg} \gg \sigma_{q\bar{q}}$ which means that charm production is almost only due to gluon interactions. In Fig. \ref{crosscc2} the cross section is plotted by varying the charm invariant mass in the range $1.1 \text{ GeV}< m_c < 1.3 \text{ GeV}$ in order to estimate the uncertainties.

\pagebreak

\section{$J/\psi$ Production}

The ALICE collaboration published some data on $c\bar{c}$ and $J/\psi$ production over the last decade \cite{alice-cc,alice-psi7,alice-psi13}. In these works, charm production was presented in terms of the relative yields by means of the normalized charged pseudorapidity and rapidity densities:

\begin{equation}
\frac{({dN}/{d\eta}(\eta=0))}{\langle {dN}/{d\eta}  \rangle}
\,\,\,\,\,\,\,\,\,\,\,\,\,\,\,\,\,\,\,\,\,\,\,\,\,\,\,\,\,\,\,\,\,\,\,\,
\frac{({dN}/{dy}(y=0))}{\langle {dN}/{dy}  \rangle}
\label{rapdens}
\end{equation}

In proton-proton collisions, the color evaporation model formula  Eq.(\ref{N cem}) can be splitted into a gluon term and a quark-antiquark term

\begin{equation}
N_{c\bar{c}}(b)=T_{pp\prime}^{gg}(b) \,  \sigma_{gg}^{CEM} + 
T_{pp\prime}^{qq}(b) \,  \sigma_{q\bar{q}}^{CEM}
\label{Ncc}
\end{equation}
This can be done because the gluon-gluon and quark-quark overlaps can be separated into Eq.(\ref{Tqq}) and (\ref{Tgg}). The gluon-quark terms are not included because these processes do not produce heavy quarks. Eq.(\ref{Ncc}) can be differentiated with respect to y as:

\begin{equation}
    \frac{dN_{c\bar{c}}}{dy}(b)=T_{pp\prime}^{gg}(b) \,    
\frac{d \sigma_{gg}^{CEM}}{dy} + T_{pp\prime}^{qq}(b) \,  
\frac{d\sigma_{q\bar{q}}^{CEM}}{dy}
    \label{dNdy}
\end{equation}

We will apply the following  the change of variables 

\begin{equation}
x_1 = \frac{p_T}{\sqrt{s}} e^y 
\,\,\,\,\,\,\,\,\,\,\,\,
x_2 = \frac{p_T}{\sqrt{s}} e^{-y}
\,\,\,\,\,\,\,\,\,\,\,\,
x_1 x_2 s = p_T^2
\,\,\,\,\,\,\,\,\,\,\,\,
dx_1 dx_2 \rightarrow \frac{2p_T}{s} dy dp_T 
\label{variables}
\end{equation}
to Eq.(\ref{sigcem}). After the change of variables $(x_1,x_2)\rightarrow (y,p_T)$, the cross section can be written as

\begin{equation}
	\sigma^{CEM} = \mathbf{F} K \sum_{i,j}\int_{(2m_c)^2}^{(2m_D)^2} dm^2  \int dy dp_T \frac{2p_T}{s} f_i(y,p_T,\mu_F^2) f_j(y,p_T,\mu_F^2) \sigma_{ij}(m^2,\mu_R^2)    \delta(m^2-p_T^2)
\label{sigcem2}
\end{equation} 

By noticing that $2p_T dp_T=dp_T^2$, and using the delta function properties to perform the integration in $p_T$, we find

\begin{equation}
	\sigma^{CEM} = \mathbf{F} K \sum_{i,j} \int_{(2m_c)^2}^{(2m_D)^2} dm^2 \int dy  \frac{1}{s} f_i(y,m,\mu_F^2) f_j(y,m,\mu_F^2) \sigma_{ij}(m^2,\mu_R^2)
\label{sigcem3}
\end{equation} 

So, the $y$ distribution can be calculated as:

\begin{equation}
	\frac{d\sigma^{CEM}}{dy} = \mathbf{F} K \sum_{i,j} \int_{(2m_c)^2}^{(2m_D)^2} dm^2    \frac{1}{s} f_i(y,m,\mu_F^2) f_j(y,m,\mu_F^2) \sigma_{ij}(m^2,\mu_R^2)
\label{sigcemdy}
\end{equation}

\begin{figure}[h!]
	\centering
    \includegraphics[scale=0.66]{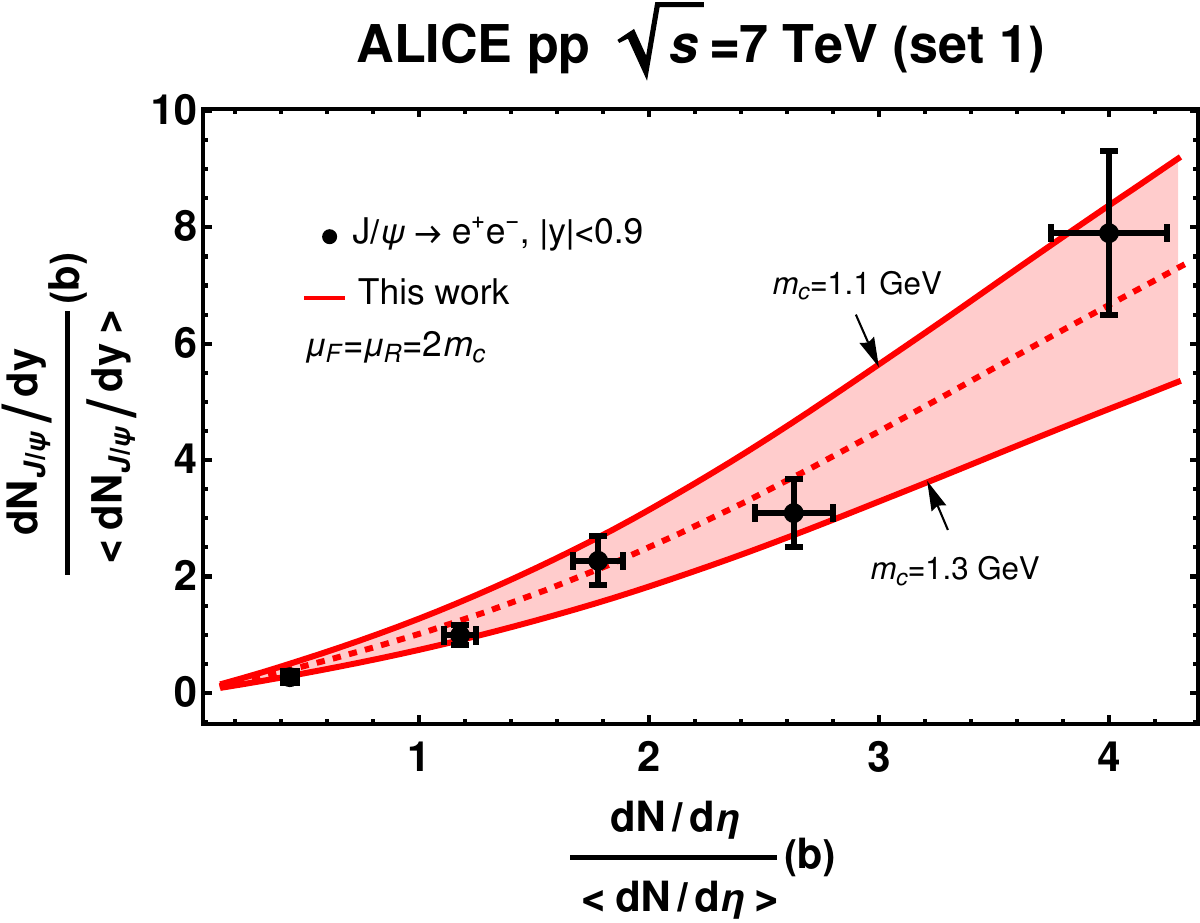}
    \caption{a) $J/\psi$ relative yield for $\sqrt{s}=7\text{ TeV}$, set 1. The experimental data are from \cite{alice-psi7}. The dashed curve is plotted with $m_c=1.2 \text{ GeV}$.}
\label{jpsi7set1}
\end{figure}

\begin{figure}[h!]
	\centering
    \includegraphics[scale=0.66]{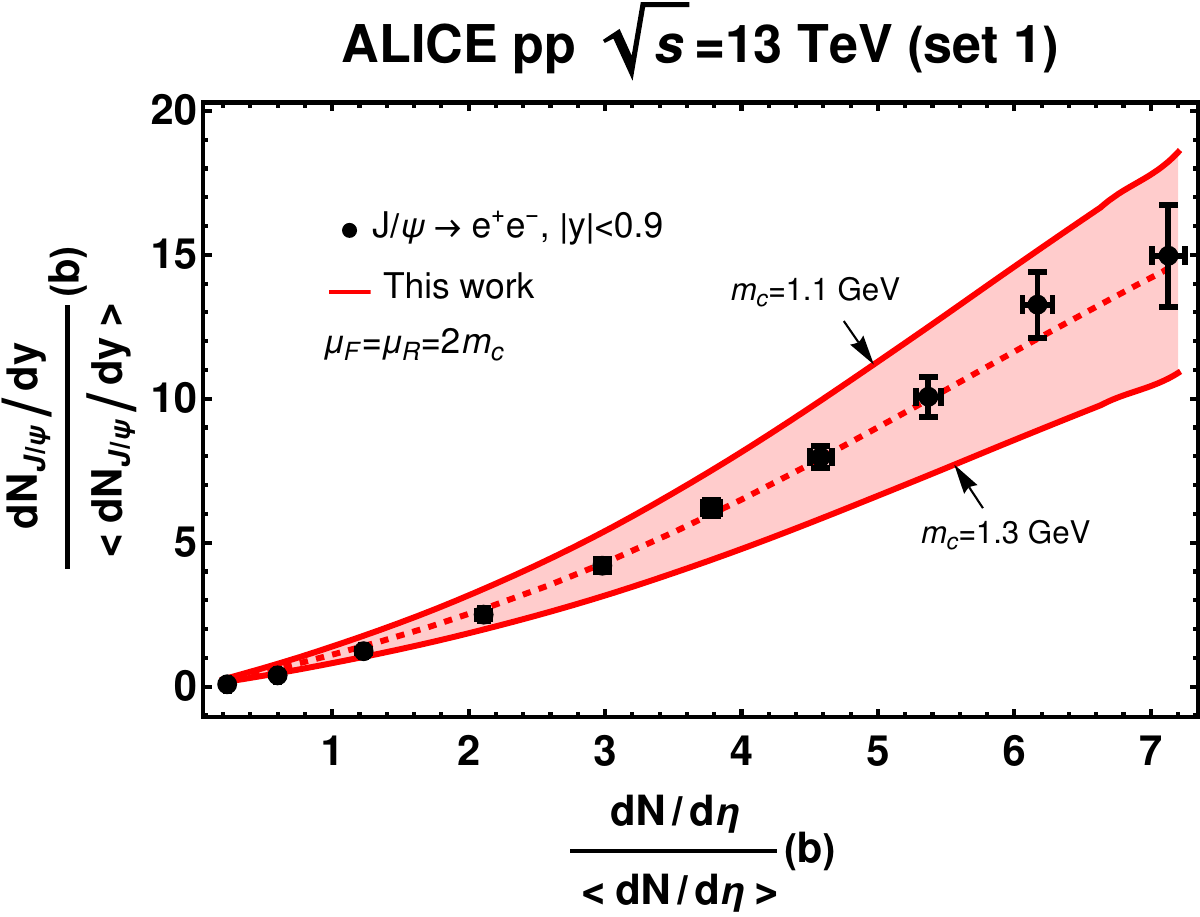}
    \caption{$J/\psi$ relative yield for $\sqrt{s}=13\text{ TeV}$, set 1. The experimental data are from \cite{alice-psi13}. The dashed curve is plotted with $m_c=1.2 \text{ GeV}$.}
\label{jpsi13set1}
\end{figure}

\begin{figure}[h!]
	\centering
    \includegraphics[scale=0.66]{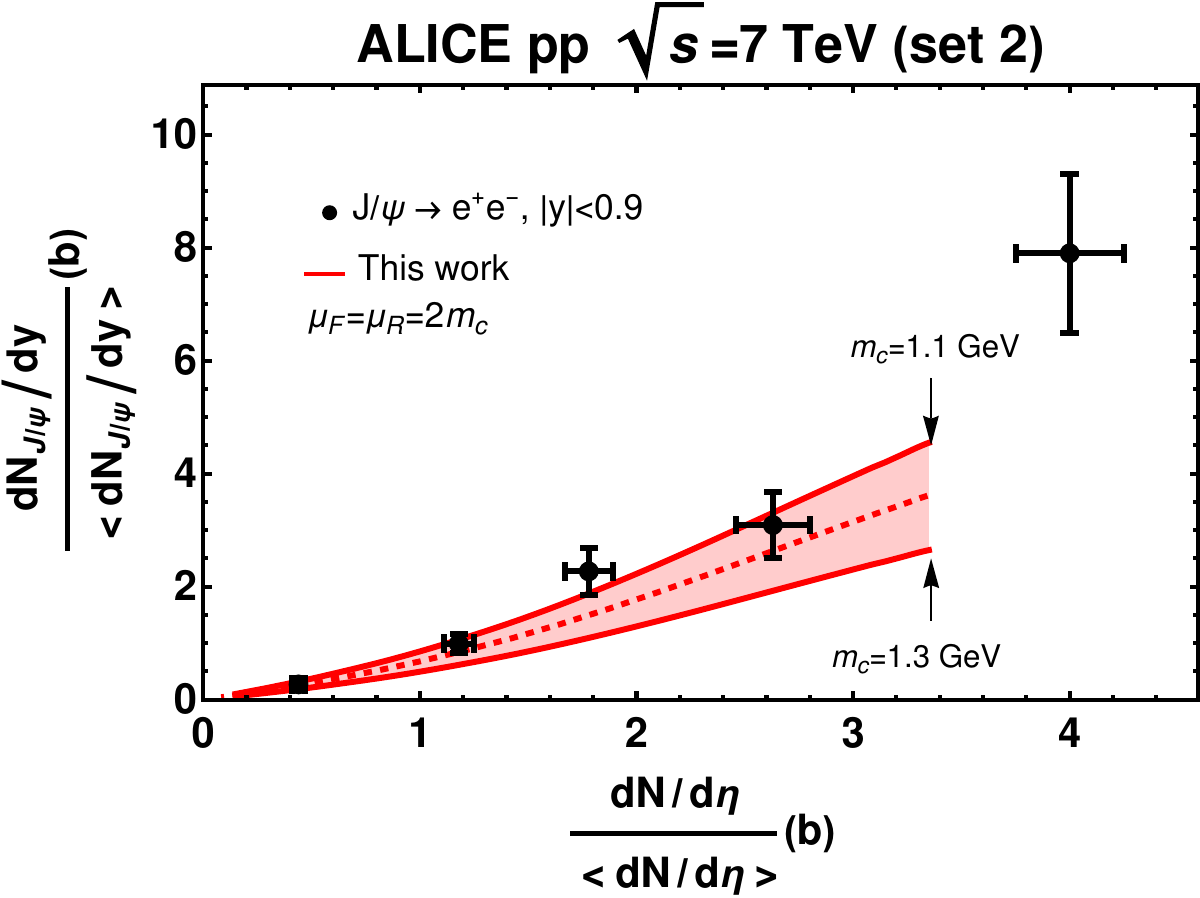}
    \caption{a) $J/\psi$ relative yield for $\sqrt{s}=7\text{ TeV}$, set 2. The experimental data are from \cite{alice-psi7}. The dashed curve is plotted with $m_c=1.2 \text{ GeV}$.}
\label{jpsi7set2}
\end{figure}

\begin{figure}[h!]
	\centering
    \includegraphics[scale=0.66]{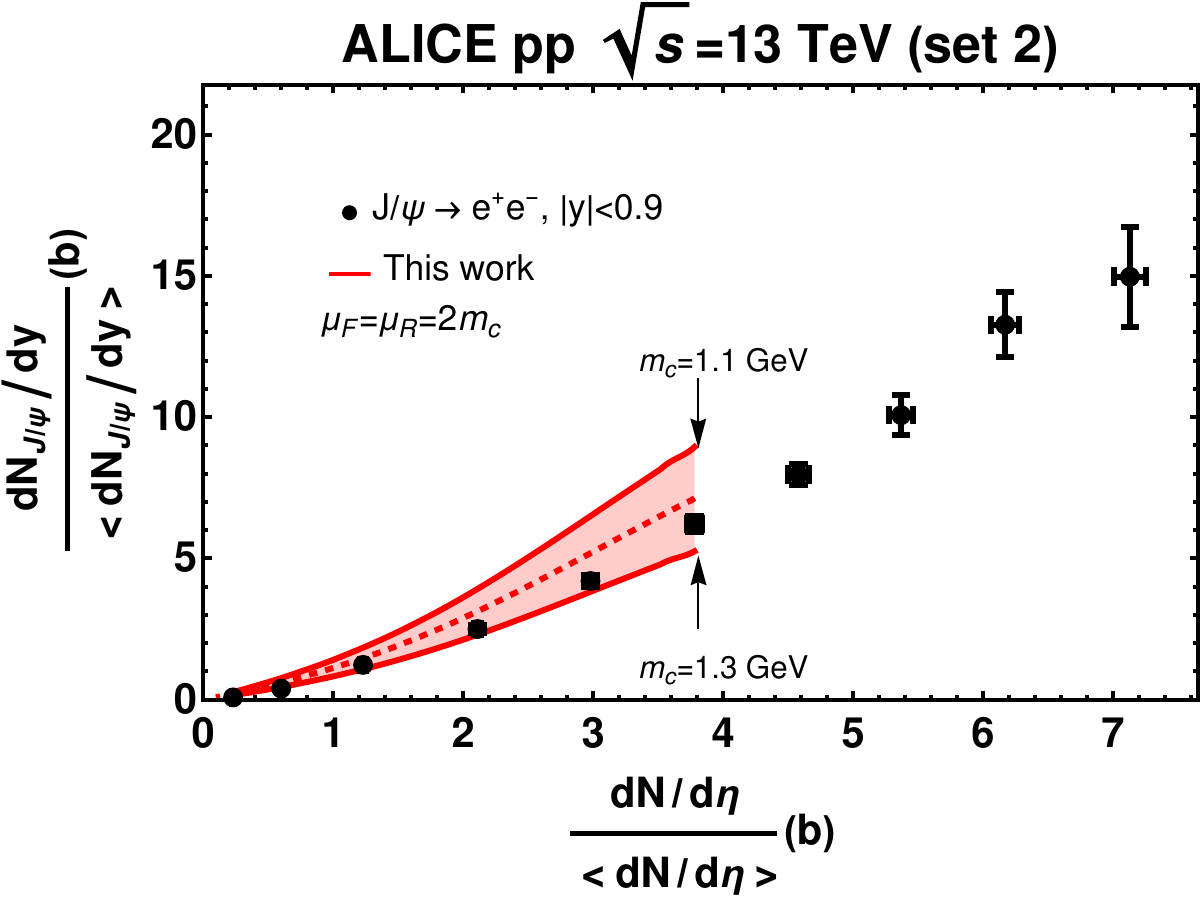}
    \caption{$J/\psi$ relative yield for $\sqrt{s}=13\text{ TeV}$, set 2. The experimental data are from \cite{alice-psi13}. The dashed curve is plotted with $m_c=1.2 \text{ GeV}$.}
\label{jpsi13set2}
\end{figure}

\begin{figure}[h!]
	\centering
    \includegraphics[scale=0.65]{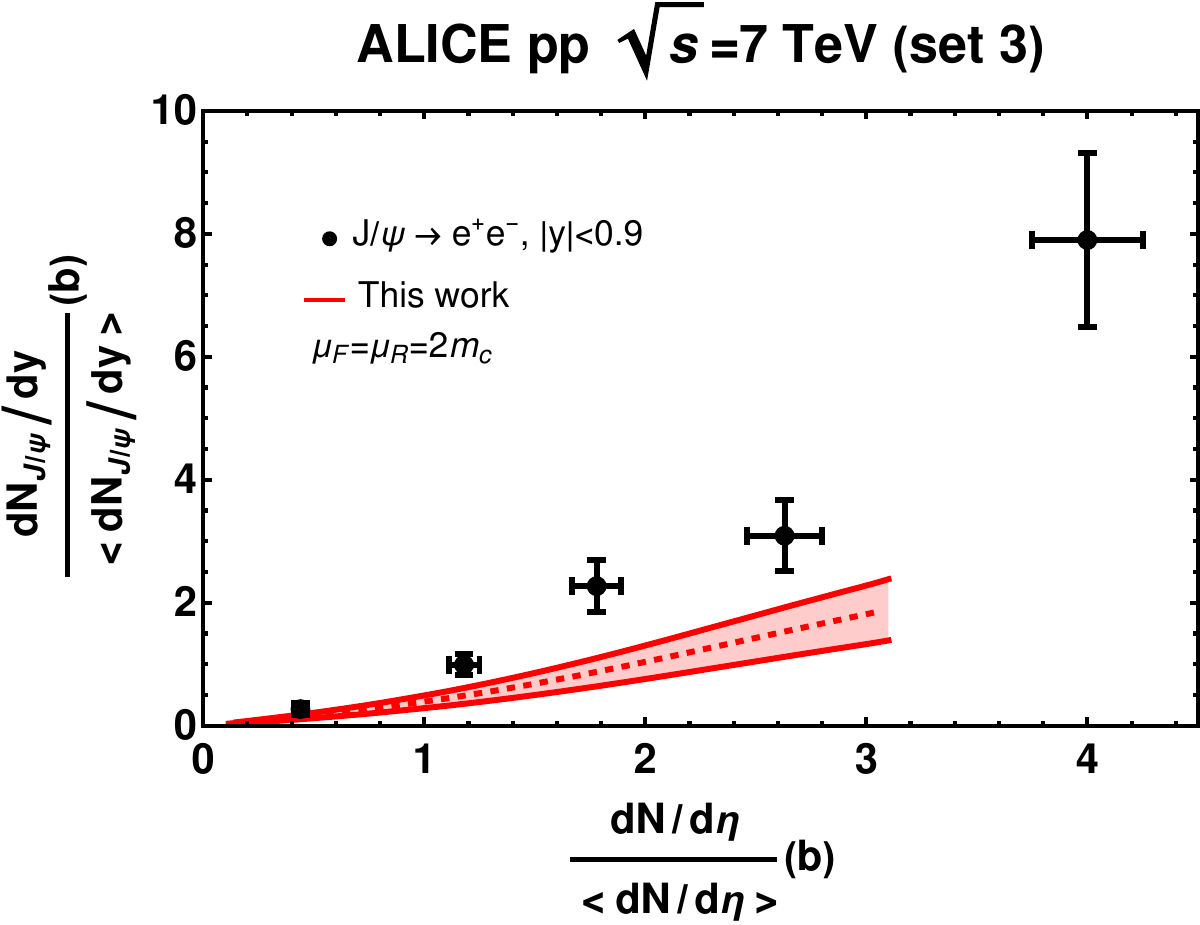}
    \caption{a) $J/\psi$ relative yield for $\sqrt{s}=7\text{ TeV}$, set 3. The experimental data are from \cite{alice-psi7}. The dashed curve is plotted with $m_c=1.2 \text{ GeV}$.}
\label{jpsi7set3}
\end{figure}

\begin{figure}[h!]
	\centering
    \includegraphics[scale=0.65]{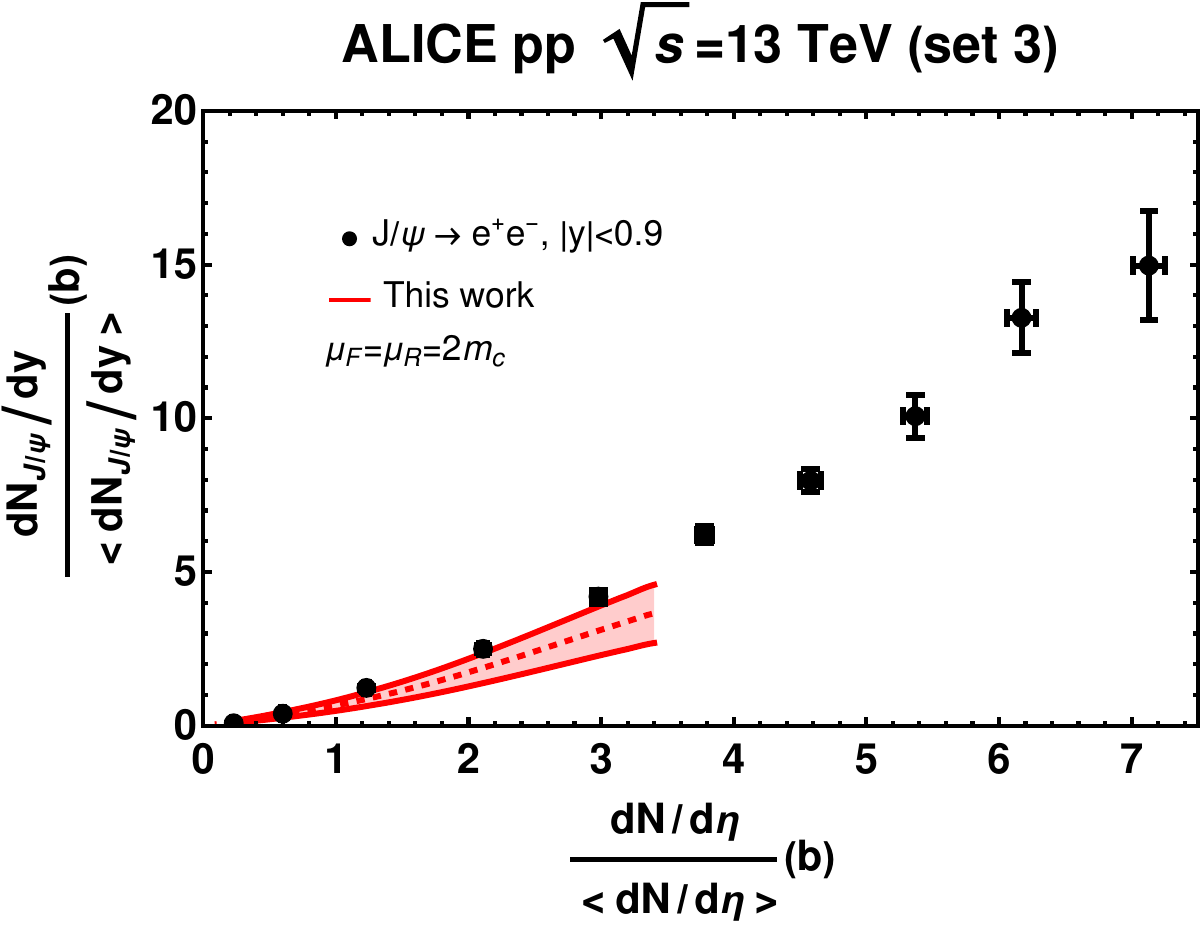}
    \caption{$J/\psi$ relative yield for $\sqrt{s}=13\text{ TeV}$, set 3. The experimental data are from \cite{alice-psi13}. The dashed curve is plotted with $m_c=1.2 \text{ GeV}$.}
\label{jpsi13set3}
\end{figure}

The central density can be calculated according to Eq.(\ref{dndeta}), with $N_{part}$ and $N_{coll}$ from the Glauber model sets of Table \ref{param-glauset}. Using the Glauber parameters of Table \ref{param-glauset} with the respective CEM parameters of Table \ref{parametrosCEM},  the experimental data for the three sets can be fitted as shown in Figs. \ref{jpsi7set1} to \ref{jpsi13set3}. The rapidity density of Eq.(\ref{sigcemdy}) is calculated in $y=0$ and, as in the open charm case, $\mu_F=\mu_R=2m_c$. The uncertainties are estimated by varying the charm mass in the range $1.1 \text{ GeV} < m_c < 1.3 \text{ GeV}$.

\begin{table}[ht!]
\caption{CEM parameter set for $pp$ collisions.}
\label{parametrosCEM}
\centering
    \begin{tabular}{| c | c | c | c |}
    \hline 
                                       & Set 1   \\ \hline
            f ($\sqrt{s} = 7$ TeV)     &  19 \%  \\ \hline
            f ($\sqrt{s} = 13$ TeV)     &  16 \%  \\ \hline
            F ($\sqrt{s} = 7$ TeV)     &  2.3 \% \\ \hline
            F ($\sqrt{s} = 13$ TeV)    &  3.4 \% \\ \hline
$ \langle dN /d \eta \rangle$ ($\sqrt{s} = 7$ TeV) & $6.0$  \cite{alice-dndeta-7tev}  \\ \hline
$ \langle dN /d \eta \rangle$ ($\sqrt{s} = 13$ TeV) & $6.4$ \cite{alice-dndeta-13tev}  \\ \hline
$ \langle dN_{J/\psi}/dy \rangle$ ($\sqrt{s} = 7$ TeV) & $8.2 \times 10^{-5}$ \cite{alice-psi7} \\ \hline
$ \langle dN_{J/\psi}/dy \rangle$ ($\sqrt{s} = 13$ TeV) & $7.9 \times 10^{-5}$ \cite{alice-psi13} \\ \hline
    \end{tabular}
\end{table}

As can be seen in Figs. \ref{jpsi7set1} and \ref{jpsi13set1}, the experimental data are better fitted with the parameter Set 1. This set presents a linear growth in the low charged multiplicity region (peripheral collisions) and it grows faster in the high charged multiplicity region (ultra-central). Following the ``core-corona" model presented in Fig.(\ref{rhos2}), the quark interactions dominate in the peripheral region, and the gluon ones dominate in the ultra-central region. Taking a look at Fig.\ref{crosscc1}, we notice that $\sigma_{gg}^{CEM}\gg \sigma_{q\bar{q}}^{CEM}$. So, as the gluon interactions become more important, the production of $J/\psi$ grows faster because of the huge difference between the cross sections. 

In Figs. \ref{jpsi7set2} and \ref{jpsi13set2}, the second parameter set fits the low charged multiplicity data (ultra-peripheral data), but not the UC data. In Fig.\ref{jpsi7set3} and \ref{jpsi13set3}, the third set is not appropriate to fit the data. It occurs because the number of partons, respectively 7 and 5, combined with the respectives parton-parton cross sections, are not sufficient to generate the high multiplicity data. In other words, the model is strongly sensitive to the degrees of freedom and to parton-parton cross sections. 

In summary, since the set with the greatest $N_g$ and smallest $\sigma^{pp}$ (Set 1) presents the best explanation for the experimental data, we can say that the collision is better explained through a larger number of point-like particles interactions.

\pagebreak

\section{$J/\psi$ Production in pA Collisions}

\begin{figure}[h!]
	\centering
    \includegraphics[scale=0.66]{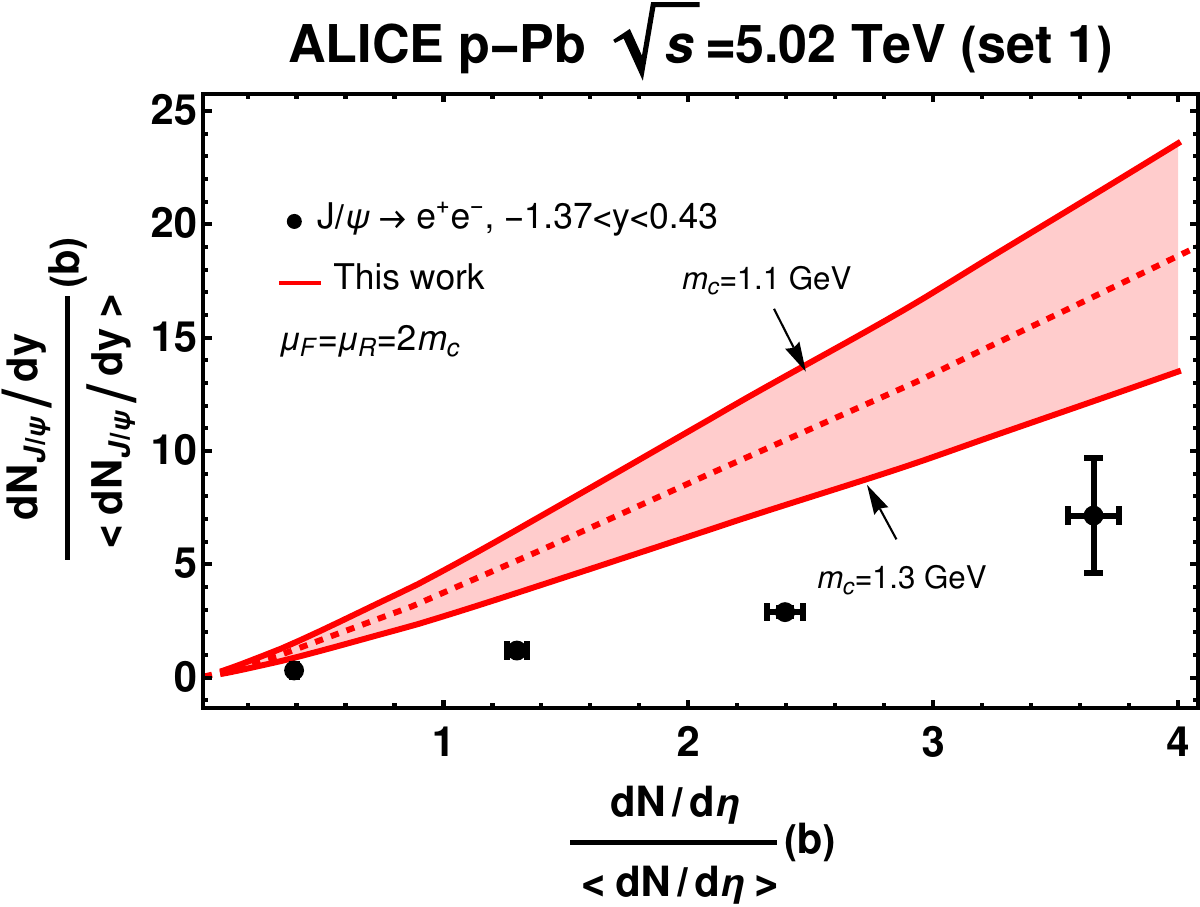}
    \caption{a) $J/\psi$ relative yield for $pPb$ at $\sqrt{s}=5.02\text{ TeV}$, set 1. The experimental data are from \cite{pPjpsi}. The dashed curve is plotted with $m_c=1.2 \text{ GeV}$.}
\label{jpsiset1pPb}
\end{figure}

\begin{figure}[h!]
	\centering
    \includegraphics[scale=0.66]{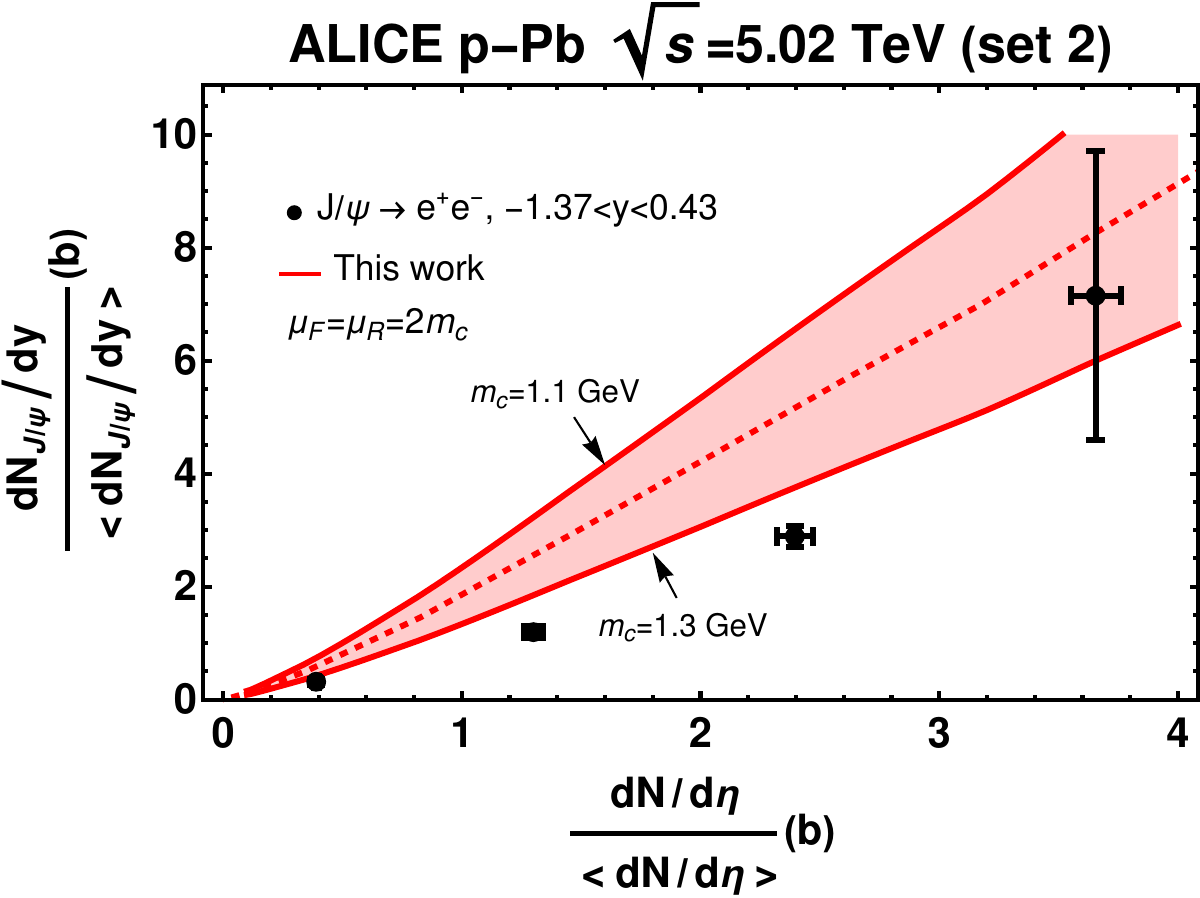}
    \caption{$J/\psi$ relative yield for $pPb$ at $\sqrt{s}=5.02\text{ TeV}$, set 2. The experimental data are from \cite{pPjpsi}. The dashed curve is plotted with $m_c=1.2 \text{ GeV}$.}
\label{jpsiset2pPb}
\end{figure}

\begin{figure}[h!]
	\centering
    \includegraphics[scale=0.66]{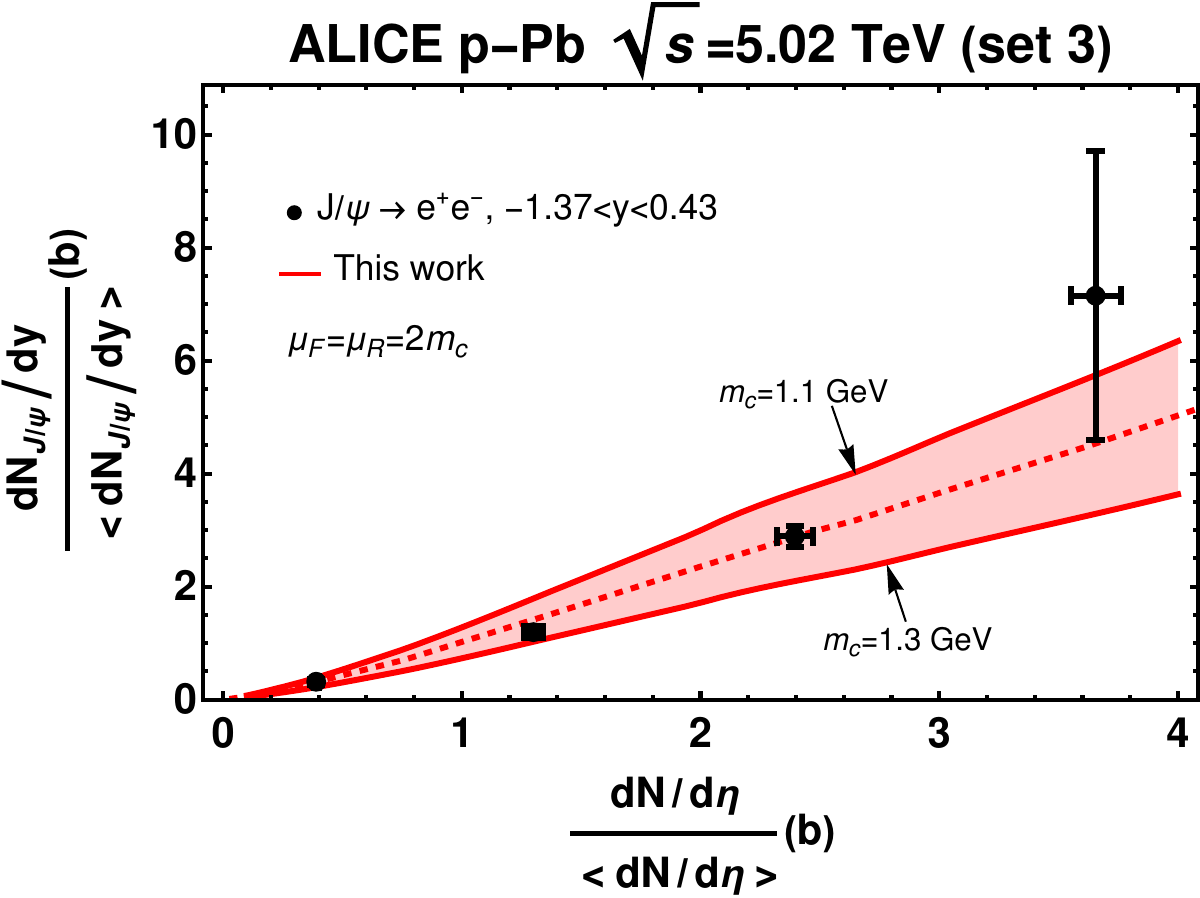}
    \caption{a) $J/\psi$ relative yield for $pPb$ at $\sqrt{s}=5.02\text{ TeV}$, set 3. The experimental data are from \cite{pPjpsi}. The dashed curve is plotted with $m_c=1.2 \text{ GeV}$.}
\label{jpsiset3pPb}
\end{figure}

Charmonium production in pA collisions can be  calculated in the same way as in $pp$ collisions. The relative yields can be calculated as:

\begin{equation}
    \frac{dN_{c\bar{c}}}{dy}(b)=T_{pPb}^{gg}(b) \,    
\frac{d \sigma_{gg}^{CEM}}{dy} + T_{pPb}^{qq}(b) \,  
\frac{d\sigma_{qq}^{CEM}}{dy}
    \label{dNdypPb}
\end{equation}
with the thickness from Eqs.(\ref{TqqpPb},\ref{TggpPb}) and the $y$ distributions of Eq.(\ref{sigcemdy}). Using the Glauber parameters of Table \ref{param-glauset} with the respective CEM of Table \ref{parametrosCEMpPb}, the results are shown from Fig. \ref{jpsiset1pPb} to \ref{jpsiset3pPb}.

\begin{table}[ht!]
\caption{CEM parameter set for $pPb$ collisions.}
\label{parametrosCEMpPb}
\centering
    \begin{tabular}{| c | c |}
    \hline 
            f ($\sqrt{s} = 5.02$ TeV)  &   20\%  \\ \hline
            F ($\sqrt{s} = 5.02$ TeV)  & 2.5\%       \\ \hline
$ \langle dN /d \eta \rangle$ ($\sqrt{s} = 5.02$ TeV) &$17.6$  \cite{pPbxMB}\\ \hline
$ \langle dN_{J/\psi}/dy \rangle$ ($\sqrt{s} = 5.02$ TeV) & $1.4 \times 10^{-4}$ \cite{pPbyMB}  \\ \hline
    \end{tabular}
\end{table}

The parameter set that better describes the experimental data is the third. Now, the enhanced growth is explained not only by the gluon distribution, but also because of the high density region of matter superposition in the center of the nucleus. As can be seen in Fig. \ref{pPbset3}, $N_{part}$ is almost constant in the UC region, but $N_{coll}$ increases substantively. In other words, the enhanced relative yields are explained by the larger number of interactions in the low impact parameter collisions.

In contrast to the $pp$ case, as shown in Figs. \ref{jpsiset1pPb} and \ref{jpsiset2pPb}, the yields are better reproduced by the smallest $N_g$ and biggest $\sigma^{pp}$. Since the curves overshoot the data for $N_g=7,10$ and $\sigma^{pp}=5.7,2.5 \text{ mb}$ it seems that the collisions are better described by a smaller number of partons, with a larger interaction range.

\pagebreak

\chapter{Conclusion}

In this work we adapted the Glauber model formalism, initially formulated to deal with heavy ion collisions, to proton-proton and proton-nucleus collisions. By doing this, we could access the subnucleonic degrees of freedom, which are important in this kind of process, and adapt a model that accounts to the baryon junction existence. This approach leads to a separation between the gluon and quark interactions and distributions.

Since there are no data about the Glauber outputs, we used the color evaporation model as a tool to test the consequences of the internal structure in the charmonium relative yields at midrapidity. In $pp$ collisions the data were described by the parameter set 1, and the enhanced growth in the high multiplicity region was explained by the core-corona model. That is, while the collision gets more central, the quark interactions are changed to gluon interactions and the relative yields grows because $gg \rightarrow c \bar{c}$ cross section is much bigger than the $q\bar{q}\rightarrow c \bar{c}$ one. In other words, the baryon junction assumption could explain the growth seen in the data for the larger number of partons.

For the $pPb$ collisions, we could adapt the nucleon structure developed for the proton by randomly positioning $208$ of them inside the nuclei. The CEM was again used to probe the internal structure by means of $J/\psi$ production and the data were fitted for the parameter set 3. Differently from the $pp$ case, the enhanced growth was not explained only by the gluon processes, but also by the enhanced number of binary collisions in the UC region.

In the end, the baryon junction assumption could explain the data. Of course our results are not enough to say that the baryon junction is completely established, but they give support to it.

\singlespacing

\cleardoublepage
\phantomsection

\end{document}